\documentclass[a4paper]{article}
\usepackage{amsmath,amssymb,theorem,fullpage,mathtools}
\usepackage[shortcuts]{extdash}
\usepackage{hyperref}
\usepackage[nosort]{cite}

\let\frac\undefined

\allowdisplaybreaks

\numberwithin{equation}{section}

\def\Maketitle{{\def\newpage{}\maketitle}}

\def\eq#1{\begin{equation}#1\end{equation}}
\long\def\subeq#1{\begin{subequations}#1\end{subequations}}

\def\Align#1{\begin{align}#1\end{align}}

\def\Aligned#1{\begin{aligned}#1\end{aligned}}
\def\Gather#1{\begin{gather}#1\end{gather}}
\def\Gathered#1{\begin{gathered}#1\end{gathered}}
\def\Multline#1{\begin{multline}#1\end{multline}}

\def\Cases#1{\begin{cases}#1\end{cases}}
\def\?{\notag}
\def\d{\partial}
\def\bd{\bar\partial}

\def\SVir{{\it SVir}}
\def\Vir{{\it Vir}}

\def\Re{\mathop{\rm Re}\nolimits}
\def\Im{\mathop{\rm Im}\nolimits}
\def\Ker{\mathop{\rm Ker}\nolimits}

\def\Res{\mathop{\rm Res}}
\def\Arg{\mathop{\rm Arg}}
\def\const{\mathop{\rm const}\nolimits}
\def\cA{{\cal A}}

\def\cF{{\cal F}}
\def\bcF{{\cal\bar F}}

\def\cH{{\cal H}}
\def\bcH{{\cal\bar H}}

\def\cL{{\cal L}}
\def\bcL{{\cal\bar L}}

\def\cM{{\cal M}}
\def\bcM{{\cal\bar M}}

\def\cN{{\cal N}}
\def\bcN{{\bar{\cal N}}}

\def\cO{{\cal O}}

\def\ve{\varepsilon}
\def\sh{\mathop{\rm sh}\nolimits}
\def\ch{\mathop{\rm ch}\nolimits}
\def\th{\mathop{\rm th}\nolimits}

\def\tg{\mathop{\rm tg}\nolimits}

\def\lcolon{\mathopen{\,:\,}}
\def\rcolon{\mathclose{\,:\,}}

\def\Z{{\mathbb{Z}}}

\def\bfa{{\bf a}}
\def\bbfa{{\bf\bar a}}

\def\boldxi{{\boldsymbol\xi}}
\def\boldeta{{\boldsymbol\eta}}
\def\ft{{\mathfrak t}}

\def\e{{\rm e}}
\def\i{{\rm i}}

\def\bz{{\bar z}}

\def\bc{{\bar c}}
\def\bh{{\bar h}}

\makeatletter

\def\section{\@startsection{section}{1}{\z@}%
                                   {-3.5ex \@plus -1ex \@minus -.2ex}%
                                   {2.3ex \@plus.2ex}%
                                   {\normalfont\normalsize\bfseries}}
\def\subsection{\@startsection{subsection}{2}{\z@}%
                                     {-3.25ex\@plus -1ex \@minus -.2ex}%
                                     {1.5ex \@plus .2ex}%
                                     {\normalfont\normalsize\bfseries\itshape}}
\def\@seccntformat#1{\csname the#1\endcsname.~~}
\long\def\@makecaption#1#2{%
  \vskip\abovecaptionskip
  \sbox\@tempboxa{\small#1. #2}%
  \ifdim \wd\@tempboxa >0.9\hsize
  {\leftskip=0.05\hsize\rightskip=0.05\hsize\relax\small
    #1. #2\par}
  \else
    \global \@minipagefalse
    \hb@xt@\hsize{\hfil\box\@tempboxa\hfil}%
  \fi
  \vskip\belowcaptionskip}
\def\Appendix{\appendix
  \def\@seccntformat##1{Appendix~\csname the##1\endcsname.~~}}

\let\over\@@over
\let\atop\@@atop
\let\above\@@above
\let\overwithdelims\@@overwithdelims
\let\atopwithdelims\@@atopwithdelims
\let\abovewithdelims\@@abovewithdelims

\makeatother

\newtheorem{theorem}{Theorem}
\newtheorem{conjecture}{Conjecture}

\begin{document}

\title{Form factors in sinh- and sine-Gordon models, deformed Virasoro algebra, Macdonald polynomials and resonance identities}
\author{Michael Lashkevich and Yaroslav Pugai,\\[\medskipamount]
\parbox[t]{0.9\textwidth}{\normalsize\it\raggedright
Landau Institute for Theoretical Physics,
142432 Chernogolovka of Moscow Region, Russia\medspace%
\footnote{Mailing address.}
Moscow Institute of Physics and Technology, 141707 Dolgoprudny of Moscow Region, Russia}
}
\date{}

\Maketitle

\begin{abstract}

We continue the study of form factors of descendant operators in the sinh- and sine-Gordon models in the framework of the algebraic construction proposed in~\cite{Feigin:2008hs}. We find the algebraic construction to be related to a particular limit of the tensor product of the deformed Virasoro algebra and a suitably chosen Heisenberg algebra. To analyze the space of local operators in the framework of the form factor formalism we introduce screening operators and construct singular and cosingular vectors in the Fock spaces related to the free field realization of the obtained algebra. We show that the singular vectors are expressed in terms of the degenerate Macdonald polynomials with rectangular partitions. We study the matrix elements that contain a singular vector in one chirality and a cosingular vector in the other chirality and find them to lead to the resonance identities already known in the conformal perturbation theory. Besides, we give a new derivation of the equation of motion in the sinh-Gordon theory, and a new representation for conserved currents.

\end{abstract}

\section{Introduction}

The main task of the relativistic quantum field theory is the calculation of correlation functions of local operators. There are two classes of models, where this task can be fulfilled exactly and analytically, and both of them are two-dimensional. First, in a class of models of conformal field theory the correlation functions are defined uniquely by an infinite-dimensional symmetry algebra and the associativity of the operator algebra\cite{Belavin:1984vu}. Second, the models that are equivalent to free bosons or fermions, like the Ising field theory\cite{Wu:1975mw} can be treated by the technique of isomonodromic deformations of linear differential equations\cite{Sato:1978ef}. The correlation functions of all other theories are only treated by means of approximate or numeric methods.

In the case of integrable massive two-dimensional models, correlation functions can be obtained by the numeric interpolation between two expansions. The short range expansion is given by the conformal perturbation theory\cite{Zamolodchikov:1990bk}, while the long range expansion is obtained by using the form factor approach\cite{Smirnov:1992vz}. In many models, for some interesting from the physical point of view operators the validity regions of both expansions overlap and the correlation functions can be found with a high precision\cite{Zamolodchikov:1990bk,Belavin:2003pu,Fateev:2009kp}.

The main difficulty of the method is caused by the fact that the operators in both approaches are defined in different ways. In the conformal perturbation theory they are defined by their conformal properties. In the simplest case, where the conformal theory is a free boson theory, they can be expressed in terms of the basic field. In the form factor approach operators are defined as sets of form factors, which are their matrix elements with respect to the eigenvectors of the Hamiltonian. The form factors are obtained exactly as solutions to a set of bootstrap equations, or form factor axioms\cite{Karowski:1978vz,Smirnov:1984sx,Smirnov:1992vz}, as soon as the exact $S$-matrix is known. The space of solutions to the form factor axioms was conjectured to be in a one-to-one correspondence with the space of local operators defined perturbatively. This conjecture has been checked for several models by some operator counting arguments\cite{Koubek:1994di,Babelon:1996xq,Jimbo:2003ge}. Though for many interesting models the form factor axioms have been solved in a general form, it is not an easy task to identify the operators in both approaches explicitly. Generally, the identification problem has not been solved. The only exception is the Ising model\cite{Cardy:1990pc}. Besides, some promising results were recently obtained for the sine-Gordon model by using the scaling limit from the lattice six-vertex model in~\cite{Jimbo:2011gv,Jimbo:2011bc}.

In this paper we concentrate on a considerably simple case of the sinh-Gordon model. There is only one scalar particle associated with the basic field $\varphi(x)$ in this case, so that the explicit expressions for form factors were obtained long ago\cite{Koubek:1993ke,Lukyanov:1997bp,Babujian:2002fi}. The full sets of form factors are known for the exponential operators $\e^{\alpha\varphi(x)}$ and the components of the energy\--momentum tensor. For the fields obtained as commutators of the exponential fields with the integrals of motion the form factors are generally known up to normalization factors. The problem of identification of the so\-/called descendant operators, i.e.\ operators containing arbitrary products of the space\--time derivatives of the field $\varphi$ and an exponential field, with solutions to the form factor axioms has not yet been solved. Here we make some steps towards this identification following the free field realization for form factors proposed in~\cite{Feigin:2008hs}.

The operators of the sinh-Gordon model are known to satisfy the so\-/called reflection relations\cite{Zamolodchikov:1995aa,Fateev:1997nn,Fateev:1998xb}. Namely, the operators $\e^{\alpha\varphi}$, $\e^{(\pm Q-\alpha)\varphi}$, $\e^{(\pm2Q+\alpha)\varphi}$ etc.\ (with some particular value $Q$) are proportional to each other. These relations can be raised to the descendant operators over these exponential fields. In~\cite{Feigin:2008hs} relations between solutions to the form factor axioms were found, which correspond to the reflection relations. Nevertheless, the reflection relations are insufficient to identify the descendant operators. Here we try to find some additional relations. We study the particular values of $\alpha$, where the reflection relations are degenerate. To do it we show that the algebraic construction underlying the free field realization of~\cite{Feigin:2008hs} is, in a sense, a certain limit of the deformed Virasoro algebra\cite{Shiraishi:1995rp} extended according to\cite{Feigin:2007arXiv0705.0427F}. Then we construct the so\-/called screening operators that map between representation of this algebra corresponding to different values of~$\alpha$. This results in some identities between operators, which can be physically interpreted as resonance identities.

The resonance identities\cite{Zamolodchikov:1990bk} appear in the conformal perturbation theory as poles in correlation functions being considered as functions of the conformal dimensions of the fields. The poles correspond to the renormalization thresholds of the operators. The residue of a pole produced by an operator $O(x)$, which demands an operator $O'(x)$ to be added for the finiteness of the resulting correlation function, is identically equal to the correlation function that contains the operator~$O'(x)$ in the place of~$O(x)$. It has already been noticed that the resonance phenomenon manifests itself in the bootstrap form factor theory as identities between form factors\cite{Belavin:2005xg,Fateev:2009kp,Lashkevich:2011ne}, and is important in practical calculations. Usually, it is rather difficult to find out the identities, and is hardly possible to prove them in the form factor framework. Here in terms of the form factors, we obtain explicitly a set of resonance identities related to even level descendant operators. Besides, we propose a simple derivation of the sinh-Gordon equation of motion (which is also a kind of resonance identity in the quantum case) and a set of higher conservation laws obtained earlier in a different form in~\cite{Babujian:2002fi}.

As a byproduct we show the so\-/called singular vectors in the Fock spaces to be expressed in terms of the Macdonald polynomials with rectangular partitions, which provides an integral representation of the corresponding polynomials for $t=-q$ (in the standard notation~\cite{Macdonald:SFHP}). For the form factor construction the case $|q|=1$ is relevant, so that the resulting integral representation appears to be convergent on the unit circle.

A part of the results presented here has been already reported without derivation, but with some extra physical explanations, in~\cite{Lashkevich:2013mca}.

The paper is organized as follows. In Section~\ref{sec:ffca}, after reminding the definition of the models under consideration and their basic properties, we describe the free field representation proposed in~\cite{Feigin:2008hs} with some development from~\cite{Alekseev:2009ik}. We also define the algebra generated by the currents $t(z)$ and $s(z)$ and review its elementary properties. In Section~\ref{sec:sdva} we consider a dressing of the deformed Virasoro algebra by a free boson. In contrast to the deformed Virasoro algebra itself, the product has a definite limit as $t\to-q$. The structure of the Fock spaces as representations of the current algebra is described in Section~\ref{sec:screenings}. We introduce the screening operators and show how to define the singular and cosingular vectors. In Section~\ref{sec:Macdonald} we show that the singular vectors can be described in terms of the Macdonald polynomials. Section~\ref{sec:phys} provides three examples of identities, which can be easily proved by means of the screening operators: the even level resonances, the equation of motion, and the conservation laws.

\section{Form factors and current algebra}
\label{sec:ffca}

In this section we recall some physical background of the construction and fix notation. The sinh-Gordon model is defined by the action
\eq{
S_{\rm shG}[\varphi]=\int d^2x\,\left({(\d_\nu\varphi)^2\over16\pi}-2\mu\ch b\varphi\right).
\label{shG-action}
}
Here $\mu$ is a coupling constant of dimension $(\text{mass})^{2+2b^2}$. Since it is the only dimensional parameter, its value is not essential: we can render it to any value by a scaling transformation. The parameter $b$ is the only dimensionless parameter, which determines the behavior of the model. It is easy to see that it can be considered as a square root of the Plank constant in the context of quantum field theory. We will think of $b$ as of a positive real number. Some of our results are applicable in the cases of purely imaginary $b$ (the sine-Gordon model) and $|b|=1$ (the roaming trajectories model), which will be discussed later.

For the simplicity of the expressions it will be convenient to use the following combinations:%
\footnote{These parameters are related to the parameters $p$, $\omega$ of~\cite{Feigin:2008hs} as follows: $r=p+1$, $q=\e^{-\i\pi}\omega^{-1}$.}
\eq{
Q=b^{-1}+b,
\qquad
r=(bQ)^{-1},
\qquad
q=\e^{-\i\pi r}.
\label{params-def}
}
Besides, we will use the light-cone variables $z=x^1-x^0$, $\bz=x^1+x^0$ in the Minkowski space, which can be considered as well as the complex coordinates $z=x^1+\i x^2$, $\bz=x^1-\i x^2$ in the Euclidean space. Correspondingly, we will use the derivatives $\d=\d/\d z$, $\bd=\d/\d\bz$.

The spectrum of the sinh-Gordon model consists of the only particle of mass $m$ such that $\mu\sim m^{2+2b^2}$. The proportionality coefficient, which depends on $b$, is known exactly\cite{Zamolodchikov:1995xk}. The scattering is purely elastic and the $S$ matrix factorizes into two-particle ones. The $S$ matrix of two particles with the rapidities $\theta_1$, $\theta_2$ depends on the difference $\theta=\theta_1-\theta_2$ and reads
\eq{
S(\theta)={\th{1\over2}(\theta-\i\pi(1-r))\over\th{1\over 2}(\theta+\i\pi(1-r))}.
\label{S-sinh}
}

Note that the sinh-Gordon theory may be considered as a perturbation of the Liouville theory
\eq{
S_{\rm L}[\varphi]=\int d^2x\,\left({(\d_\mu\varphi)^2\over16\pi}-\mu\e^{b\varphi}\right).
\label{Liouv-action}
}
The Liouville theory is a conformal field theory. Its conformal symmetry is described by the Virasoro algebra with the central charge $c=1+6Q^2$. The perturbation leading to the sinh-Gordon model is given by the term $-{\mu\over2}\int d^2x\,\e^{-b\varphi}$. Both the sinh-Gordon and the Liouville theory can be considered as perturbations of the free massless boson theory
\eq{
S_{\rm FB}[\varphi]=\int d^2x\,{(\d_\mu\varphi)^2\over16\pi}.
\label{FB-action}
}
The perturbation terms in both perturbation theories (from the Liouville theory and the free boson theory) are relevant. It means that the free boson theory is a good starting point for classifying local operators.

The space of local operators of the free boson theory consists of the exponential operators $\e^{\alpha\varphi}$ and their Fock descendants. We conjecture the same structure in the sinh-Gordon theory. It will be convenient to write the exponential operators as follows%
\footnote{The parameter $a$ will be convenient in the current algebra defined later. The definition of the parameter $a$ in terms of $\alpha$ here differs from that of~\cite{Feigin:2008hs} by the sign. This definition is physically equivalent to the old one due to the reflection property~(\ref{GV-reflection}).}
\eq{
V_a(x)=\e^{\alpha\varphi(x)},
\qquad
a={1\over2}-{\alpha\over Q}.
\label{Va-def}
}
In the sinh-Gordon theory the exponential operators $V_a$ generally have nonzero vacuum expectation values, which are known exactly~\cite{Lukyanov:1996jj}. We will denote them as
\eq{
G_a=\langle V_a(x)\rangle.
\label{Ga-def}
}
Exact expressions for the vacuum expectation values is of fundamental importance: the relate the ultraviolet and infrared asymptotics of correlation functions.

The space of local operators has a natural structure of an infinite sum of Fock spaces. Let $O(x)$ be a local operator. Since the sinh-Gordon model is a relevant perturbation of the massless free field theory, we expect the Heisenberg algebra to act on local operators in the same way as in the free field case. Define the operators $\bfa_k,\bbfa_k$ ($k\in\Z$, $k\ne0$) on the space of local operators by the relations
\eq{
(\bfa_kO)(x)=\i\oint{dz\over2\pi\i}\,(z'-z)^k\d\varphi(x')O(x),
\qquad
(\bbfa_kO)(x)=\i\oint{d\bar z\over2\pi\i}\,(\bz'-\bz)^k\bd\varphi(x')O(x).
\label{localHA-def}
}
The integrations are taken over small circles in the Euclidean space around the point~$x$. The circles are supposed to be so small that the theory in the vicinity of~$x$ is conformal, which is the free massless field theory in this case. Hence, these operators satisfy the relations
\eq{
[\bfa_k,\bfa_l]=2k\delta_{k+l,0},
\qquad
[\bbfa_k,\bbfa_l]=2k\delta_{k+l,0},
\qquad
[\bfa_k,\bbfa_l]=0.
\label{localHA-commut}
}
It can be seen that
\eq{
(\bfa_kV_a)(x)=(\bbfa_kV_a)(x)=0,
\quad k>0,
\label{anV=0}
}
so that the exponential operators $V_a$ play the role of the Fock vacuums. The creation operators $\bfa_{-k}$, $\bbfa_{-k}$ generate the Fock modules of the Heisenberg algebra~(\ref{localHA-commut}). Namely, the Fock modules are spanned on the operators
\eq{
\bfa_{-k_1}\cdots\bfa_{-k_m}\bbfa_{-l_1}\cdots\bbfa_{-l_n}V_a
={\d^{k_1}\varphi\cdots\d^{k_m}\varphi\,\bd^{l_1}\varphi\cdots\bd^{l_n}\varphi\,V_a
\over\prod^m_{i=1}(k_i-1)!\prod^n_{i=1}(l_i-1)!},
\label{descend-def}
}
which we call descendant operators over~$V_a$. The numbers $L=\sum k_i$ and $\bar L=\sum l_i$ are called the right and left levels of the descendant, providing two gradings of the Fock space. The difference $s=L-\bar L$ is the (Lorentz) spin of the operator. We will also call the pair $(L,\bar L)$ the level of a descendant operator.

This Fock module contains two naturally graded subspaces $\cH_a=\bigoplus^\infty_{L=0}(\cH_a)_L$ and $\bcH_a=\bigoplus^\infty_{\bar L=0}(\bcH_a)_{\bar L}$ called the right and the left chiral subspaces correspondingly. The right chiral subspace is generated by the operators $\bfa_{-k}$ only so that $\bar L=0$, while the left one is generated by $\bbfa_{-k}$ only so that $L=0$. This Fock module naturally factorizes into the tensor product of these subspaces, $\cH_a\otimes\bcH_a$. The whole space of local operators can be written down as
\eq{
\cH=\bigoplus_a\cH_a\otimes\bcH_a.
\label{localopspace-def}
}
In the free boson theory and in the Liouville theory, which are conformal models, the level $(L,\bar L)$ is well\-/defined, while in the sinh-Gordon theory it is not so. We define the operators (\ref{descend-def}) as the result of the perturbation from the corresponding operators in the conformal field theory, but renormalization may admix some other operators of lower dimensions and the same spin. Hence, the level is only well\--defined in the chiral sectors, where $L=s$ or $\bar L=-s$, and there is no operators from the same module to admix.

The sinh-Gordon model is a rather unusual perturbation of the free boson. This, in particular, manifests in the fact that not all the exponential operators are independent. They are related by the so\-/called reflection relations\cite{Zamolodchikov:1995aa,Fateev:1997nn,Fateev:1998xb}:
\eq{
G_{a'}V_a(x)=G_aV_{a'}(x),
\quad\text{if either $a'-a\in2\Z$ or $a'+a\in2\Z$.}
\label{GV-reflection}
}
The operators are all different on the strip $0<\Re a<1$ together with its half-boundaries $\Re a=0$, $\Im a\ge0$ and $\Re a=1$, $\Im a<0$. Every operator off the strip is obtained by reflections from some operator on the strip. These reflection relations are caused by the reflection from the exponentially growing walls of the sinh-Gordon potential. The reflection relations naturally loose their sense in the free massive boson limit $b\to0$, $\mu\simeq m^2/16\pi b^2$, where $Q$ tends to infinity.

The reflection relations (\ref{GV-reflection}) must be necessarily extended to the descendant operators producing the reflection maps between the Fock spaces $\cH_a\otimes\bcH_a$ and $\cH_{a'}\otimes\bcH_{a'}$ subject to the same condition for $a$ and~$a'$. In fact, the reflection maps factorize into the right and left components and preserve the levels. The elementary reflection maps $a\to-a$ ($\alpha\to Q-\alpha$) and $a\to2-a$ ($\alpha\to-Q-\alpha$) diagonalize in appropriate Virasoro bases~\cite{Zamolodchikov:1995aa}.

Though the results of this work are most relevant to the sinh-Gordon model, they can be also applied to the sine-Gordon model. The sine-Gordon model has the same action (\ref{shG-action}), but with an imaginary `square root of Plank constant'~$b$ and a negative coupling constant~$\mu$. This difference between the two models is rather essential. The sinh-Gordon model is a model of one neutral particle, while the spectrum of the sine-Gordon model consists of a pair of charged solitons (kink and antikink) and a series of neutral breathers. The vacuum expectation values $G_a$ of the exponential operators are different analytic functions in both cases. Nevertheless, a striking fact is that the $S$ matrix and the mass of the sinh-Gordon particle can be analytically continued to those of the first (lightest) breather of the sine-Gordon model. The same is true for the `relative' form factors, i.~e.\ the form factors divided by the VEV~$G_a$. In this paper we will only discuss the first breather relative form factors in the sine-Gordon model, which are no different from the relative form factors in the sinh-Gordon model. For the sine-Gordon model, the reflection relations (\ref{GV-reflection}) are only valid in the breather sector, while they get broken for the soliton form factors.

Note that all our results are relevant to the `roaming trajectories' model~\cite{Zamolodchikov:2006pc} as well, which is the model with $|b|=1$, $0\le\Arg b\le\pi/2$. This curve continuously connects the sinh- and sine-Gordon models. Its spectrum analytically continues that of the sinh-Gordon model, while the convergence of the form factor expansion worsens as $b$ approaches $\i$, where the Berezinskii\--Kosterlitz\--Thouless transition point of the sine-Gordon model resides.

In the paper we study the form factors of local operators. Any operator $O(x)$ can be defined by the complete set of its form factors, i.~e.\ matrix elements in the basis of the stationary states. According to~\cite{Smirnov:1992vz} the form factors of a local operator in a relativistic integrable model on the plane with the only neutral scalar particle can be expressed in terms of analytic functions $F_O(\theta_1,\ldots,\theta_N)$ of particle rapidities~$\theta_i$:
\eq{
\langle\theta_{k+1},\ldots,\theta_N|O(0)|\theta_1,\ldots,\theta_k\rangle
=F_O(\theta_1,\ldots,\theta_k,
\theta_{k+1}+\i\pi,\ldots,\theta_N+\i\pi),
\label{ff-def}
}
if $\theta_1<\theta_2<\cdots<\theta_k$, $\theta_{k+1}<\cdots<\theta_N$. The functions $F_O$ satisfy a set of linear difference equations, the form factor axioms~\cite{Karowski:1978vz,Smirnov:1984sx,Smirnov:1992vz}. Any solution to these equations define a local operator. The space of solutions is usually assumed (though not proved) to coincide with the space of local operators.

The form factors of the exponential operators $V_a(x)$ in the sinh-Gordon model were first found in~\cite{Koubek:1993ke} and then rederived in a more simple and complete (with the correct vacuum expectation values) form in~\cite{Lukyanov:1997bp} on the basis of the free field prescription for the sine-Gordon model~\cite{Lukyanov:1993pn}. The last result was generalized to the descendant operators in~\cite{Babujian:2002fi}. It is important to notice that this generalization does not mean a possibility to obtain form factors of any given descendant operator of the form~(\ref{descend-def}). It only provides a general solution to the form factor equations, and this solution was conjectured to cover the whole set of descendants of a given exponential operator. Up to now there is no general way to identify every solution to the form factors equations with some particular descendant operators. Here we will follow the construction of~\cite{Feigin:2008hs}, which uses a kind of a free field representation for the `rational part' (it will be specified later what it means) of form factors. An advantage of this construction is that for generic values of $a$ it gives a basis in the space of descendant operators. This fact was established by some counting argument. Another advantage is that it made it possible to prove some important properties of the form factors of descendant operators like the cluster property and reflection equations. Later we will find it to provide some other nontrivial identities between form factors.

Let us recall the construction of~\cite{Feigin:2008hs}. Consider the Heisenberg algebra with the generators $\hat a$, $d^\pm_n$, $n\ne0$, and the commutation relations
\eq{
[d^\pm_k,d^\pm_l]=[d^\pm_k,\hat a]=0,
\qquad
[d^\pm_k,d^\mp_l]=kA^\pm_k\delta_{k+l,0},
\label{dpm-def}
}
where
\eq{
A^\pm_k=(\pm)^k(q^{k/2}-q^{-k/2})(q^{k/2}-(-)^kq^{-k/2}).
\label{Apm-def}
}
The vacuum vectors ${}_a\langle1|$, $|1\rangle_a$ are defined as follows:
\eq{
\Aligned{
&{}_a\langle1|\hat a={}_a\langle1|a,
\qquad
\hat a|1\rangle_a=a|1\rangle_a,
\\
&{}_a\langle1|d^\pm_{-n}=0,
\qquad
d^\pm_n|1\rangle_a=0
\qquad
(n>0).
}\label{vac-def}
}
The Fock modules generated from the bra ${}_a\langle1|$ and ket $|1\rangle_a$ vacuums by the action of the Heisenberg algebra will be denoted as $\cF_a$ and $\bcF_a$ correspondingly.

Now define the vertex operators
\eq{
\lambda_\pm(x)=\exp\sum_{k\ne0}{d^\pm_kz^{-k}\over k}.
\label{lambdapm-def}
}
Their operator products look like
\eq{
\Aligned{
\lambda_\pm(z_1)\lambda_\pm(z_2)
&=\lcolon\lambda_\pm(z_1)\lambda_\pm(z_2)\rcolon,
\\
\lambda_+(z_1)\lambda_-(z_2)
&=\lambda_-(z_2)\lambda_+(z_1)=f\left(z_2\over z_1\right)\lcolon\lambda_+(z_1)\lambda_-(z_2)\rcolon
\quad(z_2\ne z_1,-z_1),
}\label{lambda-norm}
}
where
\eq{
f(z)={(z+q)(z-q^{-1})\over z^2-1}.
\label{f-def}
}
Define the currents
\eq{
t(z)=\e^{\i\pi\hat a}\lambda_-(z)+\e^{-\i\pi\hat a}\lambda_+(z),
\qquad
s(z)=\lcolon\lambda_-(z)\lambda_+(-z)\rcolon.
\label{ts-def}
}
These currents form an algebra with bilinear relations. More precisely, the generators of the algebra are the Laurent components  $t_k$, $s_k$ of the currents themselves:
\eq{
t(z)=\sum_{k\in\Z}t_kz^{-k},
\qquad
s(z)=\sum_{k\in\Z}s_kz^{-k}.
\label{tksk-def}
}
The relations can be written in terms of the currents $t(z)$,~$s(z)$:
\subeq{
\label{ts-relations}
\Align{
t(z')t(z)-t(z)t(z')
&=-C_0(s(z)-s(-z))\delta\left(-{z\over z'}\right),
\label{tt-commut}
\\
f^{-1}\left(z\over z'\right)t(z')s(z)
&=f^{-1}\left(z\over z'\right)s(z)t(z'),
\label{ts-commut}
\\
f^{-1}\left(z\over z'\right)f^{-1}\left(z'\over z\right)s(z')s(z)
&=f^{-1}\left(z\over z'\right)f^{-1}\left(z'\over z\right)s(z)s(z'),
\label{ss-commut}
}}
Here $C_0=-{1\over2}(q-q^{-1})=\i\sin\pi r$ and the delta-function is defined as $\delta(z)=\sum_{k\in\Z}z^k$.

Equation (\ref{tt-commut}) means that the product $t(z')t(z)$ has a simple pole:
\eq{
t(z')t(z)={C_0z\over z'+z}(s(z)-s(-z))+O(1)
\quad\text{as $z'\to-z$.}
\label{tt-pole}
}
According to equation (\ref{ts-commut}) the operators $t(z')$ and $s(z)$ commute at regular points and the product $t(z')s(z)$ has simple poles at $z'=\pm z$. Similarly, due to (\ref{ss-commut}) the operators $s(z')$ and $s(z)$ commute at regular points and the product $s(z')s(z)$ has double poles at $z'=\pm z$. Beside the poles, the equations (\ref{ts-relations}b,c) fix the zeroes:
\eq{
t(z')s(z)=0,
\quad\text{if $z'=qz,-q^{-1}z$;}
\qquad
s(z')s(z)=0,
\quad\text{if $z'=\pm qz,\pm q^{-1}z$.}
\label{ts-zero}
}
The model parameter $r$ enters the algebra via the combination $q=\e^{-\i\pi r}$ in the zero conditions~(\ref{ts-zero}). The coefficient $C_0$ in relation (\ref{tt-commut}) is not essential from the algebraic point of view. It only fixes the normalization of~$s(z)$. It is natural to define the highest weight vectors $\langle\delta,\sigma|$, $|\delta,\sigma\rangle$ by the conditions
$$
\Aligned{
\langle\delta,\sigma|t_{-k}=\langle\delta,\sigma|s_{-k}
&=0
\quad(k>0),
&\quad
\langle\delta,\sigma|t_0
&=\delta\langle\delta,\sigma|,
&\quad
\langle\delta,\sigma|s_0
&=\sigma\langle\delta,\sigma|,
\\
t_k|\delta,\sigma\rangle=s_k|\delta,\sigma\rangle
&=0
\quad(k>0),
&\quad
t_0|\delta,\sigma\rangle
&=\delta|\delta,\sigma\rangle,
&\quad
s_0|\delta,\sigma\rangle
&=\sigma|\delta,\sigma\rangle.
}
$$
The pair $(\delta,\sigma)$ plays the role of the highest weight. In fact, due to the scaling properties of the relation~(\ref{tt-commut}), the only value that matters is the ratio~$\delta^2/C_0\sigma$. The Fock spaces $\cF_a$, $\bcF_a$ considered as modules of the algebra (\ref{tksk-def})--(\ref{ts-relations}) are modules with the highest weight
$$
\delta=2\cos2\pi a,
\qquad
\sigma=1.
$$
We will return to a study of representations of the algebra (\ref{ts-relations}) in Section~\ref{sec:screenings}.

The current $t(z)$ is the main building block for form factors of local operators. Consider the functions
\eq{
f(\theta_1,\ldots,\theta_N)
=\rho^N\langle v|t(\e^{\theta_1})\cdots t(\e^{\theta_N})|v'\rangle\prod_{i<j}R(\theta_i-\theta_j),
\label{f(theta)-def}
}
where
\eq{
\Gathered{
\rho=\left(2\cos{\pi r\over2}\right)^{-1/2}\exp\int^{\pi(1-r)}_0{dt\over2\pi}\,{t\over\sin t},
\\
R(\theta)=\exp\left(
-4\int^\infty_0{dt\over t}\,
{\sh{\pi t\over2}\sh{\pi(1-r)t\over2}\sh{\pi rt\over2}
\over\sh^2\pi t}\ch(\pi-\i\theta)t
\right).
}\label{R-def}
}
In the simplest case, where the vectors $\langle v|$ and $|v'\rangle$ are the vacuum vectors of the Fock spaces ${}_a\langle1|$ and $|1\rangle_a$ correspondingly, the functions $f(\theta_1,\ldots,\theta_N)$ satisfy the form factor axioms and provide the form factors of the exponential operators~$G_a^{-1}V_a(x)$ obtained in~\cite{Lukyanov:1997bp}. One may expect that form factors of every descendant operator are of the same form~(\ref{f(theta)-def}) with appropriate vectors $\langle v|$ and~$|v'\rangle$. These form factors as functions of the rapidity variables $\theta_i$ naturally factorize into the common for all operators transcendent part proportional to the product $\prod_{i<j}R(\theta_i-\theta_j)$ and the part rational in the variables $x_i=\e^{\theta_i}$ given by the matrix element itself, which is specific for any particular operator. This rational part will be the main object of our study.

A natural question is what kind of restriction should be imposed on the vectors $\langle v|$, $|v'\rangle$ to guarantee that the functions $f(\theta_1,\ldots,\theta_N)$ satisfied the form factor axioms. The form factors axioms were found to be satisfied if $\langle v|D_{-k}=0$, $D_k|v'\rangle=0$ ($k>0$), where the operators $D_k$ have a simple form
\eq{
D_k=d^-_k+(-)^kd^+_k.
\label{Dk-def}
}
Hence, one may define the subspaces $\cF^\cA_a$, $\bcF^\cA_a$ according to
\eq{
\cF^\cA_a=\bigl\{\langle v|\in\cF_a\>\big|\>\langle v|D_{-k}=0,\ k>0\bigr\},
\qquad
\bcF^\cA_a=\bigl\{|v\rangle\in\bcF_a\>\big|\>D_k|v\rangle=0,\ k>0\bigr\}.
\label{cFphys-altdef}
}
In~\cite{Feigin:2008hs} these subspaces were called `physical' (in quotes). To avoid confusion with the genuine space of states we will call them here $\cA$-subspaces, and their elements $\cA$-vectors.

Notice that the current $s(z)$ is expressed in terms of the operators $D_k$ only:
\eq{
s(z)=\lcolon\exp\sum_{k\ne0}{D_kz^{-k}\over k}\rcolon.
\label{sDk-rel}
}
This makes it possible to define a $\cA$-subspace of an arbitrary module of the algebra~(\ref{ts-relations}). We will call the vector $\langle v|$ or $|v\rangle$ an $\cA$-vector, if
\eq{
\Aligned{
\langle v|s_{-k}
&=0\quad(k>0),
&\quad
\langle v|s_0
&=\sigma\langle v|,
\\
s_k|v\rangle
&=0\quad(k>0),
&\quad
s_0|v\rangle
&=\sigma|v\rangle.
}\label{physstate-bs-def}
}
It should be mentioned that the $\cA$-subspaces are not modules of the algebra~(\ref{ts-relations}).

In the free field realization (\ref{ts-def}) the $\cA$-subspaces can be constructed explicitly. Consider the commutative algebra $\cA$ generated by the elements $c_{-1},c_{-2},\ldots$. There are two representations of this algebra in the Heisenberg algebra~(\ref{dpm-def}), $\pi$ and $\bar\pi$:
\eq{
\pi(c_{-k})={d^-_k-d^+_k\over A^+_k},
\qquad
\bar\pi(c_{-k})={d^-_{-k}-d^+_{-k}\over A^+_k}.
\label{pibpi-def}
}
It is easy to check that $[\pi(c_{-k}),D_l]=[\bar\pi(c_{-k}),D_l]=0$. Hence, the vectors
\eq{
{}_a\langle h|={}_a\langle1|\pi(h),
\qquad
|h\rangle_a=\bar\pi(h)|1\rangle_a,
\qquad
h\in\cA,
\label{h-state-def}
}
are $\cA$-vectors, and all $\cA$-vectors have the form~(\ref{h-state-def}). Therefore, the $\cA$-subspaces of $\cF_a$, $\bcF_a$ consist of the vectors generated by the elements of the algebra~$\cA$:
\eq{
\cF^\cA_a=\bigl\{{}_a\langle h|\>\big|\>h\in\cA\bigr\}.
\qquad
\bcF^\cA_a=\bigl\{|h\rangle_a\>\big|\>h\in\cA\bigr\},
\label{cFphys-def}
}

Define the matrix elements between the $\cA$-vectors%
\footnote{In \cite{Feigin:2008hs} these functions were denoted as $\tilde J^{h\bar h'}_a$ (with the tilde) and the corresponding operator (see below) as~$\tilde V^{h\bar h'}_a(x)$.}
\eq{
J^{h\bh'}_a(x_1,\ldots,x_N)
={}_a\langle h|t(x_1)\cdots t(x_N)|h'\rangle_a.
\label{Jhh'-def}
}
They are symmetric functions of the variables $x_1,\ldots,x_N$. We will often use the shorthand notation
\eq{
X=(x_1,\ldots,x_N),
\qquad
t(X)=t(x_1)\cdots t(x_N),
\qquad
s(X)=s(x_1)\cdots s(x_N).
\label{shorthand}
}
Then the set of the functions
\eq{
F^{h\bh'}_a(\theta_1,\ldots,\theta_N)
=G_a\rho^NJ^{h\bh'}_a(\e^{\theta_1},\ldots,\e^{\theta_N})\prod_{i<j}R(\theta_i-\theta_j),
\qquad N=0,1,2,\ldots,
\label{Fhh'-def}
}
satisfying the form factor axioms, define an operator~$V^{h\bh'}_a(x)$.

As we have already mentioned, the operator $V^{1\bar1}_a(x)$, which corresponds to the vacuum-vacuum matrix elements, coincides with the exponential operator $V_a(x)$. The operators $V^h_a(x)\equiv V^{h\bar1}_a(x)$ and $V^\bh_a(x)\equiv V^{1\bar h}_a(x)$ are right and left chiral descendants of the exponential operator $V_a(x)$ correspondingly. In other words $V^h_a\in\cH_a$, $V^\bh_a\in\bcH_a$. For generic values of the parameter $a$ there exist the isomorphisms $\cF^\cA_a\cong\cH_a$, $\bcF^\cA_a\cong\bcH_a$ of the graded spaces: $(\cF^\cA_a)_L\cong(\cH_a)_L$, $(\bcF^\cA_a)_L\cong(\bcH_a)_L$.

Indeed, their dimensions coincide
\eq{
\dim(\cF^\cA_a)_L=\dim(\bcF^\cA_a)_L=\dim\cA_L=\dim(\cH_a)_L=\dim(\bcH_a)_L=P_L,
\label{dim-id}
}
where $P_L$ is the number of partitions of the integer~$L$. We may say that in a sense the restriction to the $\cA$-subspaces is necessary to eliminate the extra boson, which enters the definition of the $t(z)$ current. It was noted above that the level in the chiral sectors can be identified with the spin of an operator, which is well\-/defined in any relativistic quantum field theory. Besides, in the limits $a\to\pm\i\infty$ the operators $V^h_a(x)$ (and $V^\bh_a(x)$) can be proved to differ for different elements $h\in\cA$. The deformation argument proves this fact for generic values of~$a$. It means that for generic $a$ there are as many linearly independent operators $V^h_a$ on each level as the chiral descendants. This proves the existence of the isomorphism, but does not provide any hint to how to construct it. In fact, it is only known for some particular elements $h$, and mostly up to a normalization factor. Some examples will be discussed later.

In general, the operators $V^{h\bh'}_a(x)$ were conjectured to coincide with some operators from the space $\cH_a\otimes\bcH_a$. It is tempting to assert the isomorphism between $\cF_a\otimes\bcF_a$ and $\cH_a\otimes\bcH_a$ as doubly graded spaces. It seems to be true, but up to now we have too few examples, where it can be checked explicitly. The obstacle is that the total level $L+\bar L$ (in contrast to the spin $L-\bar L$) is not well\-/defined in a massive field theory. The only thing that can be said for sure is that if the elements $h$ and $h'$ are of level $L$ and $\bar L$ correspondingly, the operator $V^{h\bh'}_a$ is a linear combination of operators of levels that not exceed $(L,\bar L)$.

The equation (\ref{Fhh'-def}) reduces studying form factors to studying the functions~$J^{h\bh'}_a(X)$. These functions are easily calculated by using the relations (\ref{lambda-norm}), (\ref{f-def}) and the properties:
\subeq{\label{pibpi-props}
\Align{{}
[\pi(c_{-k}),\lambda_\pm(z)]
&=(\mp)^{k+1}z^k\lambda_\pm(z),
&\pi(c_{-k})|1\rangle_a
&=0,
\label{pilambda-commut}
\\
[\bar\pi(c_{-k}),\lambda_\pm(z)]
&=-(\pm)^{k+1}z^{-k}\lambda_\pm(z),
&{}_a\langle1|\bar\pi(c_{-k})
&=0,
\label{bpilambda-commut}
\\
[\pi(c_{-k}),\bar\pi(c_{-l})]
&=-(1+(-1)^k)k(A^+_k)^{-1}\delta_{kl}.
\label{pibpi-commut}
}}
We see the odd elements $c_{-2k+1}$ to act trivially: the action of $\pi(c_{-2k+1})$ multiplies any matrix element by $\sum^N_{i=1}x_i^{2k-1}$, while that of $\bar\pi(c_{-2k+1})$ multiplies it by $\sum^N_{i=1}x_i^{-2k+1}$. These elements realize the adjoint action of the integrals of motion $I_{\pm(2k-1)}$ of spin $\pm(2k-1)$ on local operators:
\eq{
V^{c_{-2k+1}h\bh'}_a(x)\sim[I_{2k-1},V^{h\bh'}_a(x)],
\qquad
V^{h\,\overline{c_{-2k+1}h'}}_a(x)\sim[I_{-2k+1},V^{h\bh'}_a(x)].
\label{IM-action}
}
We may say that the action of the isomorphism map $\cF^\cA\to\cH_a$ on $c_{-2k+1}$ is known up to a normalization factor. In the particular case of $c_{-1}$ it is known exactly:
\eq{
{m\over2}V^{c_{-1}h\bh'}_a(x)=[P_z,V^{h\bh'}_a(x)]=-\i\d V^{h\bh'}_a(x),
\qquad
-{m\over2}V^{h\,\overline{c_{-1}h'}}_a(x)=[P_\bz,V^{h\bh'}_a(x)]=-\i\bd V^{h\bh'}_a(x),
\label{PbarP-action}
}
where $P_z$ and $P_\bz$ are components of the momentum vector.

On the contrary, the even elements $c_{-2k}$ act nontrivially. The commutation of $\pi(c_{-2k})$ or $\bar\pi(c_{-2k})$ with $\lambda_-(z)$ and with $\lambda_+(z)$ produce factors with opposite signs, which cannot be factored out. Besides, note that the nonvanishing commutator~(\ref{pibpi-commut}) is responsible for any nonvanishing vacuum expectation values of the operators $V^{h\bh'}(x)$.

Now discuss the symmetries of the functions~$J^{h\bh'}_a$. First, there is an evident periodicity property:
\eq{
J^{h\bh'}_a(X)=(-)^NJ^{h\bh'}_{a+1}(X).
\label{Jhh'-periodicity}
}

Another invariance is related to the antiautomorphism of the construction (\ref{dpm-def})--(\ref{ts-def})
\eq{
(d_k^\pm)^\top=-d_{-k}^\mp,
\qquad
\hat a^\top=-\hat a,
\qquad
\left(d\over da\right)^\top={d\over da},
\qquad
q^\top=q,
\qquad
z^\top=z^{-1},
\label{conjugation}
}
This antiautomorphism does not conjugate complex numbers, so that it can be considered as a kind of transposition. The operator $d/da$, defined by the relation $[d/da,\hat a]=1$, will be used later. The antiautomorphism (\ref{conjugation}) induces a one-to-one linear map of the spaces
\eq{
\cF_a\stackrel\top\longleftrightarrow\bcF_{-a}.
\label{cFbcF-isomorph}
}
This invariance leads to the identity
\eq{
J^{h\bh'}_a(x_1,\ldots,x_N)=J^{h'\bh}_{-a}(1/x_1,\ldots,1/x_N).
\label{J-conj}
}
The algebraic construction is not symmetric with respect to the space reflection $x^1\to-x^1$ or, equivalently, $z\leftrightarrow-\bz$, which interchanges left and right chiral sectors. But this identity means that it is invariant under the simultaneous space reflection and the substitution $a\to-a$ ($\alpha\to Q-\alpha$). To restore the full invariance we have to use the reflection symmetry~(\ref{GV-reflection}). In the algebraic construction this symmetry is nontrivial.

Namely, it was proved\cite{Feigin:2008hs} that there exists a continuous family of linear maps $r_a:\cA\to\cA$, called the reflection maps, such that
\eq{
J^{h\bh'}_a(X)=J^{r_a(h)\,\overline{r_{-a}(h')}}_{-a}(X).
\label{Jhh'-reflection}
}
To avoid confusion it should be stressed that the map $r_a$ is not an automorphism. For the unit element it is trivial: $r_a(1)=1$. For the odd generators $c_{-2k+1}$ it is also trivial and admits factorization: $r_a(c_{-2k+1}h)=c_{-2k+1}r_a(h)$ for any $h\in\cA$. As for the elements that contain the even generators $c_{-2k}$, the reflection map is explicitly known for the element $c_{-2}$ only:
\eq{
r_a(h^{(2)}_a)=h^{(2)}_{-a},
\qquad
h^{(2)}_a={c_{-2}-\i c_{-1}^2\tg\pi a\over\sin\pi r+\sin2\pi a}.
\label{h2-reflection}
}
The property (\ref{Jhh'-reflection}) of the operators $V^{h\bh'}_a$, together with the periodicity, which follows from~(\ref{Jhh'-periodicity}), corresponds to the reflection property~(\ref{GV-reflection}) in the Lagrangian field theory and its generalization to the descendant operators.

The proof of the reflection property (\ref{Jhh'-reflection}) was based on the following two facts. First, it can be checked straightforwardly, by means of some recursion relations, that
\eq{
{}_a\langle1|t(X)s(Y)|1\rangle_a={}_{-a}\langle1|t(X)s(Y)|1\rangle_{-a}.
\label{ts-reflection}
}
Second, for a \emph{generic} value of $a$ the Fock spaces $\cF_a$ and $\bcF_a$ are irreducible modules of the algebra~(\ref{ts-relations}). Hence, the whole Fock spaces can be generated by means of the products of the form $t(\boldxi)s(\boldeta)$, $\boldxi=(\xi_1,\ldots,\xi_r)$, $\boldeta=(\eta_1,\ldots,\eta_s)$. Namely, consider the vectors
\eq{
{}_a\langle1|t(z^{-1}\boldxi^{-1})s(z^{-1}\boldeta^{-1})
=\sum^\infty_{k=0}z^k\>{}_a\langle\boldxi,\boldeta;k|,
\qquad
t(z\boldxi)s(z\boldeta)|1\rangle_a
=\sum^\infty_{k=0}z^k|\boldxi,\boldeta;k\rangle_a.
\label{XiEtak-def}
}
From the definition (\ref{cFphys-altdef}) it is easy to check that ${}_a\langle\boldxi,\boldeta;k|\in\cF^\cA_a$ and $|\boldxi,\boldeta;k\rangle_a\in\bcF^\cA_a$ for $1\le k\le L$, if
\eq{
\sum^r_{i=1}\xi_i^k+(1+(-1)^k)\sum^s_{j=1}\eta_j^k=0,
\qquad
1\le k\le L.
\label{XiEta-cond}
}
The statement~\cite{Alekseev:2009ik} is that for generic values of $a$ it is possible to choose bases in the spaces $(\cF_a)_L$ and $(\bcF_a)_L$ by using the vectors ${}_a\langle\boldxi,\boldeta;L|$ and $|\boldxi,\boldeta;L\rangle_a$ with different values of $\boldxi$, $\boldeta$ that satisfy the condition~(\ref{XiEta-cond}). It reduces the proof of the reflection property~(\ref{Jhh'-reflection}) to the first fact expressed by~(\ref{ts-reflection}).

In the present paper we will focus on the case of the particular values of~$a$, for which the vectors of the form ${}_a\langle\boldxi,\boldeta;k|$ and $|\boldxi,\boldeta;k\rangle_a$, $k=0,1,2,\ldots$, do \emph{not} span the spaces $\cF^\cA_a$,~$\bcF^\cA_a$. We may call the corresponding spaces $\cF_a$, $\bcF_a$ the degenerate modules of the algebra~(\ref{ts-relations}). The simplest case of such a particular value is $a=1/2$, which corresponds to the unit operator and its descendants. Since all form factors of the unit operator except the zero-particle one vanish, the above construction produces no vectors except the vacuum one. The operators $V^{c_{-1}}_{1/2}$, $V^{\bc_{-1}}_{1/2}$, and $V^{c_{-1}\bc_{-1}}_{1/2}$, being derivatives of the unit operator, are zero, so that all their form factors explicitly vanish. The operator $V^{c_{-2}}_{1/2}$ is (up to a factor and an additive full derivative) the energy-momentum tensor component~$T_{zz}$. For an extended discussion of this case see Section~\ref{subsec:phys:cl}.

A more complicated example, discussed in~\cite{Feigin:2008hs}, corresponds to $a=1-{r\over2},1-{1-r\over2}$. For these values of $a$ the level 2 element $h=c_{-2}-\i c_{-1}^2\tg\pi a$ is nonvanishing, while $J^h_a=0$. This identity immediately follows from (\ref{h2-reflection}), since the denominator of $h^{(2)}_a$ vanishes at the points $a=1-{r\over2},1-{1-r\over2}$, while $h^{(2)}_{-a}$ is finite at these values. It seems rather difficult to explicitly construct such `reflection invariant' families at higher levels, and up to now we are unable to generalize this proof. Here we develop an alternative approach to obtaining such `null vectors' based on the so\-/called screening operators. The screening operators commute with the currents $t(z)$, $s(z)$ and may be considered as intertwining operators between the Fock modules $\cF_a$ or $\bcF_a$ with different values of the parameter~$a$. We demonstrate that the degenerate modules appear for the values of $a$ corresponding to $\alpha=\pm\alpha_{mn}$, where
\eq{
\alpha_{mn}={1-m\over2}b^{-1}+{1-n\over2}b,
\qquad
m,n=1,2,3,\ldots
\label{alphamn-def}
}
The values $\alpha=\alpha_{mn}$ correspond to the Kac dimensions of the Virasoro algebra of the Liouville theory~(\ref{Liouv-action}). The values $\alpha=-\alpha_{mn}$ correspond to the Kac dimensions in the Liouville theory, which is obtained from (\ref{Liouv-action}) by the substitution $\varphi\to-\varphi$. Though it is not certain that there is no other degenerate points in the parameter~$a$, such conjecture seems very plausible. We will describe the structure of the Fock spaces as modules of the algebra~(\ref{ts-relations}). Before doing it let us describe the algebra as a particular case of a well-known algebra, the deformed Virasoro algebra.

\section{Symmetrized deformed Virasoro algebra and its \texorpdfstring{$t\to-q$}{t->-q} limit}
\label{sec:sdva}

The deformed Virasoro algebra~\cite{Shiraishi:1995rp} $\Vir_{q,t}$, which depends on two deformation parameters $q$ and $t$, is generated by the elements $T_n$ ($n\in\Z$) satisfying some quadratic relations. It is convenient to write down these relations in terms of the currents $T(z)=\sum_{k\in\Z}T_kz^{-k}$. Let $p=q/t$. The relations read
\eq{
\chi(z/z')T(z')T(z)-\chi(z'/z)T(z)T(z')
=-{(q^{1/2}-q^{-1/2})(t^{1/2}-t^{-1/2})\over p^{1/2}-p^{-1/2}}(\delta(pz/z')-\delta(z/pz')),
\label{Vir-def}
}
where
\eq{
\chi(z)=\exp\left(-\sum^\infty_{k=1}\chi_kz^k\right),
\qquad
\chi_k={(q^{k/2}-q^{-k/2})(t^{k/2}-t^{-k/2})\over k(p^{k/2}+p^{-k/2})}.
\label{chi-def}
}
Let us make a brief comment on the notation. In this section we do not consider the parameter $q$ as the sinh-Gordon model parameter defined in~(\ref{params-def}). It is just a deformation parameter. Its relation to the field theory will be clear later.

In this paper we are interested in the limit $p\to\e^{-\i\pi}$. We see that in this limit the algebra cannot be defined by the relations (\ref{Vir-def}) since the denominator of $\chi_k$ vanishes for odd values of~$k$. To overcome this difficulty consider the algebra $H\otimes\Vir_{q,t}$, where $H$ is the Heisenberg algebra generated by the operators $h'_k$, $k\in\Z\setminus\{0\}$ with the commutation relations
\eq{
[h'_k,h'_l]=-\chi_k\delta_{k+l,0}.
\label{h'-def}
}
Define the Fock module $\cF^{h'}$ built over the vacuum $|0\rangle^{h'}$, such that $h'_k|0\rangle^{h'}=0$, $k>0$. As usual, the normal ordering $\lcolon\cdot\rcolon$ puts every element $h'_k$ with $k>0$ to the right of~$h'_{-k}$, e.g.\ $\lcolon h'_kh'_l\rcolon=\lcolon h'_lh'_k\rcolon=h'_lh'_k$, if $k>0$.

Define the currents
\eq{
t(z)=T(z)\lcolon\e^{\sum_{k\ne0}h'_kz^{-k}}\rcolon,
\qquad
s(z)=\lcolon\e^{\sum_{k\ne0}(1+p^{-k})h'_kz^{-k}}\rcolon,
\label{ts(z)qt-def}
}
and the function
\eq{
f_{qt}(z)={(z-q^{-1})(z-t)\over(z-1)(z-p^{-1})}.
\label{fqt-def}
}
From the defining relations (\ref{Vir-def}) and (\ref{h'-def}) we obtain the following relations for the new currents:
\subeq{
\label{ts-qt-relations}
\Align{
\omit\span
t(z')t(z)-t(z)t(z')
=-{(q^{1/2}-q^{-1/2})(t^{1/2}-t^{-1/2})\over p^{1/2}-p^{-1/2}}
\left(s(z)\delta\left(pz\over z'\right)-s(z')\delta\left(z\over pz'\right)\right),
\label{tt-qt-commut}
\\
f_{qt}^{-1}\left(z\over z'\right)t(z')s(z)
&=f_{qt}^{-1}\left(z\over z'\right)s(z)t(z'),
\label{ts-qt-commut}
\\
f_{qt}^{-1}\left(z\over z'\right)f_{qt}^{-1}\left(z'\over z\right)s(z')s(z)
&=f_{qt}^{-1}\left(z\over z'\right)f_{qt}^{-1}\left(z'\over z\right)s(z)s(z').
\label{ss-qt-commut}
}}
We will refer to the subalgebra $\SVir_{q,t}$ in $H\otimes\Vir_{q,t}$ generated by the Laurent components of the currents $t(z)$, $s(z)$ as the symmetrized deformed Virasoro algebra. This algebra is a particular case of the Feigin\--Kojima\--Shiraishi\--Watanabe algebra\cite{Feigin:2007arXiv0705.0427F} (the $s=1$ case in the notation of that article). An advantage of the algebra $\SVir_{q,t}$ is that it admits a regular limit as $p\to\e^{-\i\pi}$. Since $f_{q,-q}(z)=f(z)$, the relations (\ref{ts-qt-relations}) turn to (\ref{ts-relations}) for $q=-t$. We will identify the limiting algebra $\SVir_{q,-q}$ with the algebra~(\ref{ts-relations}). To study the limit it is convenient to use the standard free field representation of the deformed Virasoro algebra\cite{Shiraishi:1995rp,Awata:1996xt}.

Let $h_k$ ($k\in\Z$), $d/dh_0$ be the generators of the Heisenberg algebra with the commutation relations
\eq{
[h_k,h_l]=\chi_k\delta_{k+l,0},
\qquad
\left[{d\over dh_0},h_0\right]=1.
\label{h-def}
}
Let $r=\log q/\log p$, $b^2={1-r\over r}=-\log t/\log q$, and $Q=b+b^{-1}$. The Fock vacuum vector $|a\rangle^h$ is defined by the conditions: $h_k|a\rangle^h=0$, $k>0$; $h_0|a\rangle^h=Q(a-1/2)|a\rangle^h$. The corresponding Fock module will be denoted by~$\cF^h_a$. Again, the normal ordering places every element $h_k$ with $k>0$ to the right of every~$h_{-k}$.

The realization of $T(z)$ in terms of the free bosons is given by
\eq{
T(z)=\Lambda_-(z)+\Lambda_+(z),
\qquad
\Lambda_\pm(z)
=p^{\pm1/2}q^{\pm bh_0}\lcolon\exp\left(\pm\sum_{k\ne0}h_kp^{\pm k/2}z^{-k}\right)\rcolon.
\label{T-ffield}
}
Evidently, the operators $\Lambda_\pm(z)$ do not admit a well\-/defined limit as $p\to\e^{-\i\pi}$ due to the zeros in the denominator in~(\ref{h-def}). Consider the linear combinations
\eq{
d^-_k=k(-p^{-k/2}h_k+h'_k),
\qquad
d^+_k=k(p^{k/2}h_k+h'_k),
\label{dpm-qt-def}
}
which satisfy the commutation relations
\eq{
[d^\pm_k,d^\pm_l]=0,
\qquad
[d^+_k,d^-_l]=kp^{k/2}(q^{k/2}-q^{-k/2})(t^{k/2}-t^{-k/2})\delta_{k+l,0}.
\label{dpm-qt-commut}
}
Reversely, the generators $h_k$, $h'_k$ are expressed in terms of the operators $d^\pm_k$ as
\eq{
h_k=-{d^-_k-d^+_k\over k(p^{k/2}+p^{-k/2})},
\qquad
h'_k={p^{k/2}d^-_k+p^{-k/2}d^+_k\over k(p^{k/2}+p^{-k/2})}.
\label{hh'-dpm}
}
Reparameterizing the zero mode operators
\eq{
\hat a={h_0\over Q}+{1\over2},
\qquad
{d\over da}=Q{d\over dh_0},
\label{hata-qt-def}
}
we obtain the following representation for the currents:
\eq{
t(z)=p^{-\hat a}\lambda_-(z)+p^{\hat a}\lambda_+(z),
\qquad
s(z)=\lcolon\lambda_-(z)\lambda_+(pz)\rcolon,
\qquad
\lambda_\pm(z)=\exp\sum_{k\ne0}{d^\pm_k\over k}z^{-k}.
\label{ts-qt-ffield}
}
Since the commutation relations (\ref{dpm-qt-commut}) turn to (\ref{dpm-def}) in the limit $p\to\e^{-\i\pi}$, the operators $t(z)$, $s(z)$ tend to those defined in the construction for form factors described in the previous section.

An important ingredient of the deformed Virasoro algebra is the screening operators. Define first the screening currents
\subeq{
\label{SvS-qt-def}
\Align{
S(z)
&=\e^{(1-r){d\over da}}
\lcolon\exp\left(-\sum_{k\ne0}{p^{k/2}+p^{-k/2}\over q^{k/2}-q^{-k/2}}h_kz^{-k}\right)\rcolon.
\label{S-qt-def}
\\
\tilde S(z)
&=\e^{r{d\over da}}
\lcolon\exp\left(\sum_{k\ne0}{p^{k/2}+p^{-k/2}\over t^{k/2}-t^{-k/2}}h_kz^{-k}\right)\rcolon,
\label{vS-qt-def}
}}
The screening currents were originally introduced in~\cite{Lukyanov:1994re} and then applied to the construction of the deformed Virasoro algebra in~\cite{Shiraishi:1995rp}. The currents defined in~\cite{Shiraishi:1995rp} and~\cite{Feigin:2007arXiv0705.0427F} are related to the currents~(\ref{SvS-qt-def}) as follows:%
\footnote{Note that the plus and minus subscripts in~\cite{Shiraishi:1995rp} are interchanged in comparison to the notation accepted in the conformal field theory.}
$$
S_+(z)=F_1(z)=S(z)z^{2\hat a-1\over r},
\qquad
S_-(z)=E_1(z)=\tilde S(z)z^{2\hat a-1\over1-r}.
$$
In fact, we will only need one of these currents, e.g.~$S(z)$. Henceforth we will only discuss the screening current $S(z)$ and the corresponding screening operator. The results below can be easily transferred to $\tilde S(z)$ by means of the substitution $q\leftrightarrow t^{-1}$, $r\to1-r$.

The screening operators are, generally, integrals of screening currents that commute with the Virasoro algebra. For $|q|<1$ they are given by Lukyanov's formula~\cite{Lukyanov:1996qs}:
\eq{
\Gathered{
X(z_0)=\oint{dz\over2\pi\i}\,S_+(z){[v-v_0+{1\over2}-2\hat a]\over[v-v_0-{1\over2}]},
\qquad
z=p^v,\ z_0=p^{v_0},
\\
[v]=p^{{1\over2}\left({v^2\over r}-v\right)}(p^v;q)_\infty(qp^{-v};q)_\infty(q;q)_\infty,
\qquad
(z;q)_\infty=\prod^\infty_{k=0}(1-zq^k).
}\label{Lukscr}
}
The elliptic function $[\cdot]$ possesses the properties
\eq{
\textstyle
[v+r]=-[v]=[-v],
\qquad
[v+{2\pi\i\over\log p}]=-\e^{{2\pi\i\over r}\left(v+{\i\pi\over\log p}\right)}[v].
\label{ellbracket-props}
}
It was proved~\cite{Jimbo:1996vu} that on the space
\eq{
\cF^h_{mn}=\cF^h_{a_{mn}},
\label{bcFmn-def}
}
where
\eq{
a_{mn}={r\over2}m+{1-r\over2}n,
\label{amn-def}
}
the $n$th power of the screening operator commutes with the deformed Virasoro algebra:
\eq{
[X^n(z_0),T(z)]|_{\cF^h_{mn}}=0.
\label{XnT-commut}
}

In this paper we are interested in the case $|q|=|p|=1$, where we encounter a difficulty. It is easy to see that for odd values of~$n$ Lukyanov's screening operator can be defined in the limit $p\to\e^{-\i\pi}$. Indeed, by using the properties (\ref{ellbracket-props}) we obtain
\Multline{
{[v+{1\over2}-2a_{mn}]\over[v-{1\over2}]}
={[v+{1\over2}+r(n-m)-n]\over[v-{1\over2}]}
=(-)^{n-m}{[v+{1\over2}-n]\over[v-{1\over2}]}
\to\const\times\e^{-\i\pi(n-1)v/r}
\\
\quad\text{as $p\to\e^{-\i\pi}$, if $n\in2\Z+1$.}
\?
}
In this case the screening operator $X(z_0)$ is $z_0$-independent and is proportional to $\oint{dz\over2\pi\i}z^{m-n}S(z)$. But it is possible to show that $X^s=0$, $s>1$, while the powers play an important role in the representation theory of the deformed Virasoro algebra. For even $n$ it is even worse: the operators $X^s$ contain divergences on the unit circle $|q|=1$ becoming completely senseless. Thus we need an alternative definition. We propose such a definition in the next section. We introduce an analog of the power $X^s$ for the particular case $p=\e^{-\i\pi}$, which is relevant for form factors, for the values of $s$, such that $n-s$ is an even integer.

\section{Action of screening operators on Fock spaces}
\label{sec:screenings}

It was proved in~\cite{Feigin:2008hs} that for generic values of $a$ every operator $V^{h\bh'}_a(x)$ is nonzero, if $h,h'\ne0$. It means the existence of such an integer $N$ and a set $X=(x_1,\ldots,x_N)$ that ${}_a\langle h|t(X)|h'\rangle_a\ne0$. Here we are interested in the particular values of the parameter $a$ where there exist such elements $h,h'\in\cA$, for which $V^{h\bh'}_a(x)=0$. This takes place for the values~$a=a_{mn}$ defined in~(\ref{amn-def}). These values correspond to $\alpha=\alpha_{mn}$ in the sense that
\eq{
V_{mn}(x)\equiv V_{a_{mn}}(x)=\e^{\alpha_{mn}\varphi(x)}.
\label{Vmn-def}
}
The operators $V_{mn}$ for $m,n>0$ are primary fields corresponding to the degenerate modules of the Virasoro algebra of the Liouville theory~(\ref{Liouv-action}). Note that the operators $V_{2-m,2-n}$ for $m,n>0$ correspond to the same degenerate modules of the second Virasoro algebra, obtained by the substitution $\varphi\to-\varphi$, as it has been discussed at the end of Section~\ref{sec:ffca}.

Even for a generic $r$ the operators $V_{mn}$ are not quite independent. Indeed, the reflection transformations identify some operators with different values of $m$ and~$n$. For the values $a=a_{mn}$ the equation (\ref{GV-reflection}) takes the form
\eq{
G_{m'n'}V_{mn}=G_{mn}V_{m'n'},
\quad\text{if $m-m',n-n'\in4\Z$ or $m+m',n+n'\in4\Z$,}
\label{amn-reflsim}
}
where $G_{mn}=G_{a_{mn}}$. Besides, we should remember that the quasiperiodicity (\ref{Jhh'-periodicity}) means a relation between form factors of the operators $V^{h\bh'}_{mn}\equiv V^{h\bh'}_{a_{mn}}$ and~$V^{h\bh'}_{m+2,n+2}$. Note that for some values of $m$ and $n$ the vacuum expectation values $G_{mn}$ vanish. In the sinh-Gordon model it results in vanishing the corresponding operators:
\eq{
V_{mn}(x)=V_{2-m,2-n}(x)=0,\quad\text{$m\ge n\ge2$, $n\in2\Z$ or $n\ge m\ge2$, $m\in2\Z$.}
\label{Vmn-zero}
}
Below we will understand the product $G_{mn}^{-1}V_{mn}(x)$ as a limit of $G_a^{-1}V_a(x)$, which is always finite.

On the contrary, in the sine-Gordon model all operators $V_{mn}(x)$ are nonzero due to the fact that the normalization condition is well\-/defined for all values of~$a$. Though some vacuum expectation values vanish in this case, the corresponding soliton form factors remain nonzero. Moreover, the form factors possess poles as functions of $a$ due to the resonance phenomenon in the sine-Gordon model.

Introduce the notation for the Fock spaces of the Heisenberg algebra (\ref{dpm-def}):
\eq{
\cF_{mn}=\cF_{a_{mn}},
\qquad
\bcF_{mn}=\bcF_{a_{mn}}.
\label{cFbcFmn-def}
}
The spaces $\cF_{mn}$ and $\bcF_{mn}$ are dual to each other and, hence, their structures as modules of the algebra $\SVir_{q,-q}$ are different. Due to the bijection (\ref{cFbcF-isomorph}), the structure of the Fock space $\cF_{mn}$ coincides with that of~$\bcF_{-m,-n}$.

The corresponding vectors are defined as
\eq{
{}_{mn}\langle v|
={}_{a_{mn}}\langle v|\in\cF_{mn},
\qquad
|v\rangle_{mn}
=|v\rangle_{a_{mn}}\in\bcF_{mn}.
\label{vec-mn-def}
}
The matrix elements of the form $J^{h\bar h'}_{mn}(X)={}_{mn}\langle h|t(X)|h'\rangle_{mn}$ determine the form factors $F^{h\bh'}_{mn}$ of the operators $V^{h\bar h'}_{mn}(x)=V^{h\bar h'}_{a_{mn}}(x)$.

Our next goal is to define the so\-/called screening operators $Q^{(s)}$, which commute with the algebra $\SVir_{q,-q}$ and intertwine between different modules, e.~g.\ $\cF_a$ and~$\cF_{a'}$. We will find this commutativity requirement to restrict the values of $a$, $a'$ to the special values of the form~$a_{mn}$. First, notice that the screening current (\ref{S-qt-def}) in the case $t=-q$ looks like
\eq{
S(z)=\delta\lcolon\exp\sum_{k\ne0}{d^-_k-d^+_k\over k(q^{k/2}-q^{-k/2})}z^{-k}\rcolon,
\qquad
\delta=\e^{(1-r){d\over da}}.
\label{S-def}
}
The operator $\delta$ only shifts the value of $a$ and maps the Fock space vectors identically: $\delta|v\rangle_a=|v\rangle_{a+r-1}$ and, hence, $\delta|v\rangle_{mn}=|v\rangle_{m,n-2}$. Due to its presence the screening current changes the value of~$a$ too: $S(z):\bcF_a\to\bcF_{a+r-1}$. For special Fock modules it reads as $S(z):\bcF_{mn}\to\bcF_{m,n-2}$.

In what follows we will need operator products of the screening current with other operators. The products with $\lambda_\pm(z)$ can be easily computed:
\eq{
S(z')\lambda_\pm(z)=-q^{\pm1}\lambda_\pm(z)S(z')
={z'+q^{\pm1/2}z\over z'-q^{\mp1/2}z}\lcolon S(z')\lambda_\pm(z)\rcolon.
\label{Slambda-prod}
}
Each of these two products contains a single pole with the same value of the residue:
\eq{
\Res_{z=q^{\mp1/2}x}S(z)\lambda_\pm(x)=B_1x\sigma(x),
\qquad
B_1=q^{1/2}+q^{-1/2}=2\cos{\pi r\over2}.
\label{ResSlambda}
}
The operator $\sigma(z)$ here is defined as
\eq{
\sigma(z)=\lcolon S(q^{\mp1/2}z)\lambda_\pm(z)\rcolon
=\delta\lcolon\exp\sum_{k\ne0}{q^{k/2}d^-_k-q^{-k/2}d^+_k\over k(q^{k/2}-q^{-k/2})}z^{-k}\rcolon.
\label{sigma-def}
}

The product with the current $s(z)$ is trivial:
\eq{
S(z')s(z)=s(z)S(z')=\lcolon S(z')s(z)\rcolon.
\label{Ss-prod}
}

At last, the products of two screening currents and a product of the screening current with the operator $\sigma(z)$ are
\Align{
S(z')S(z)
&=-\left(z\over z'\right)^2S(z)S(z')
=\left(1-{z^2\over z^{\prime2}}\right)\lcolon S(z')S(z)\rcolon.
\label{SS-prod}
\\
S(z')\sigma(z)
&={z^2\over z^{\prime2}}\sigma(z)S(z')
={(z'+q^{1/2}z)(z'+q^{-1/2}z)\over z^{\prime2}}\lcolon S(z')\sigma(z)\rcolon.
\label{Ssigma-prod}
}
Note that there are no poles off zero in these two products.

It is convenient to rewrite the commutation relations in terms of the modes $S_k:(\bcF_a)_l\to(\bcF_{a+r-1})_{l-k}$ of the screening current~$S(z)$:
\eq{
S(z)=\sum_{k\in\Z}S_kz^{-k},
\qquad
S_k=\oint{dz\over2\pi\i}\,z^{k-1}S(z).
\label{Sk-def}
}
We obtain
\eq{
S_kS_{k'}=-S_{k'+2}S_{k-2},
\quad
[S_k,t(x)]=x^k\sigma(x)\gamma_k(\hat a),
\quad
[S_k,s(x)]=0,
\quad
S_k\sigma(x)=x^2\sigma(x)S_{k-2},
\label{Sk-commut}
}
where
\eq{
\gamma_k(a)=2B_1\cos\left(\pi a-{\pi r\over2}(k-1)\right).
\label{fk-def}
}
We will see later that the operators $S_k$, due to their nice commutation relations, are very convenient to produce vectors in the Fock spaces.

Now we are ready to introduce the screening operators. There are two basic blocks: the single screening operator $\Sigma$ and the double one~$W$:
\Align{
\Sigma|_{\bcF_{mn}}
&=\oint{dz\over2\pi\i}\,z^{m-n}S(z)=S_{m-n+1},
\label{Sigma-def}
\\*
W|_{\bcF_{mn}}
&=\oint{dz_1\over2\pi\i}\oint{dz_2\over2\pi\i}\,
z_1^{m-n+2}z_2^{m-n}S(z_1)S(z_2)F^n(z_2/z_1)
=\sum^\infty_{k=1}F^n_kS_{m-n+3-k}S_{m-n+1+k},
\label{W-def}
}
where $F^n(z)$ is a formal series of the form
\eq{
F^n(z)=\sum^\infty_{k=1}F^n_kz^k=\sum^\infty_{k=1}(-)^{k-1}{q^{k/2}-(-)^nq^{-k/2}\over q^{k/2}+(-)^nq^{-k/2}}z^k.
\label{Fn-def}
}
First of all, the action of the operator $W$ is well\-/defined on any finite level vector ${}_{mn}\langle v|$ or~$|v\rangle_{mn}$. Indeed, the operators in the r.h.s.\ of (\ref{W-def}) are ordered in such a way that for large enough $k$ the first operator in the product annihilates the vector ${}_{mn}\langle v|$ on the left, while the second one annihilates $|v\rangle_{mn}$ on the right. It means that, in fact, a finite number of terms only contribute the sum on a finite level vector.

Evidently, the operators $\Sigma$ and $W$ change the grading according to the rule
\eq{
\Sigma:(\bcF_{mn})_l\to(\bcF_{m,n-2})_{l+n-m-1},
\qquad
W:(\bcF_{mn})_l\to(\bcF_{m,n-4})_{l+2n-2m-4}.
\label{SigmaW-grad}
}
Besides, from eq.~(\ref{Sk-commut}) it immediately follows that
\eq{
\Sigma^2=0,
\qquad
[\Sigma,W]=0.
\label{SigmaW-commut}
}
The commutators $[\Sigma,t(x)]$, $[W,t(x)]$ arise from the poles in the operator products~(\ref{Slambda-prod}). The poles in both the terms of $t(x)$ lead to the same operator $\sigma(x)$ at the same point, but with different coefficients. A simple calculation gives
\eq{
\Aligned{{}
[\Sigma,t(x)]|_{\bcF_{mn}}
&=\i^n(1+(-1)^n)B_1x^{m-n+1}\sigma(x),
\\
[W,t(x)]|_{\bcF_{mn}}
&=\i^n(1-(-1)^n)B_1x^{m-n+1}\Sigma\sigma(x).
}\label{SigmaW-t-commut}
}
Hence, for even $n$ the operators $W^s$ commute with $t(x)$, while for odd $n$ the operators $\Sigma W^s$ do. Besides, the operators $\Sigma$ and $W$ commute with $s(x)$ and~$\sigma(x)$. This motivates us to define the operators
\eq{
Q^{(s)}:(\bcF_{mn})_l\to(\bcF_{m,n-2s})_{l-s(m-n+s)}
\quad\text{for $s=n\bmod2$}
\label{Q-gen-action}
}
in terms of the operators $\Sigma$ and $W$:
\eq{
Q^{(s)}=\Cases{W^{s/2},&\text{if $s\in2\Z$;}\\
\Sigma W^{(s-1)/2},&\text{if $s\in2\Z+1$.}
}\label{Q-gen:def}
}
From (\ref{SigmaW-t-commut}) we immediately obtain
\eq{
[Q^{(s)},t(x)]|_{\bcF_{mn}}=0.
\label{Qt-commut}
}
The condition $s=n\bmod2$ is important and will be assumed everywhere below except several expressions, where it will be denoted as~$\overline Q^{(s)}$ to stress the difference. Besides, it is not difficult to check that the commutator $[Q^{(s)},t(x)]$, for $Q^{(s)}$ defined with $m,n$ corresponding to the module $\bcF_{mn}$, is nonzero on any module $\bcF_a$ off the point $a=a_{mn}$.

There are two more commutation relations
\Gather{
[Q^{(s)},s(x)]=0
\quad\Leftrightarrow\quad
[Q^{(s)},D_k]=0,
\label{QsDk-commut}
\\
[Q^{(s)},\sigma(x)]=0.
\label{Qsigma-commut}
}
Here the operators $D_k$ are those defined in~(\ref{Dk-def}). According to~(\ref{cFphys-altdef}) they define the $\cA$-subspaces $\cF^\cA_a$,~$\bcF^\cA_a$. Hence, the commutation relation (\ref{QsDk-commut}) means that the action of the operators $Q^{(s)}$ is well\-/defined on the $\cA$-subspaces:
\eq{
Q^{(s)}:(\bcF^\cA_{mn})_l\to(\bcF^\cA_{m,n-2s})_{l-s(m-n+s)},
\label{Q-gen-phys-action}
}

Now let us study the action of the $Q^{(s)}$ operators on the Fock spaces. First of all, from (\ref{Q-gen-action}) we immediately obtain
\eq{
Q^{(s)}|1\rangle_{mn}=0,
\quad
{}_{-m,-n}\langle1|Q^{(s)}=0,
\quad\text{if $n<m+s$;}
\label{Q1-zero}
}
Besides, for generic value of $r$ we have
\eq{
{}_{mn}\langle N^{(s)}|\equiv{}_{m,n-2s}\langle1|Q^{(s)}\ne0,
\quad
|N^{(s)}\rangle_{-m,-n}\equiv Q^{(s)}|1\rangle_{-m,-n+2s}\ne0,
\quad
\text{if $n\le m+s$.}
\label{Nvec-def}
}
We call these level $s(m-n+s)$ vectors in the spaces $\cF^\cA_{mn}$ and $\bcF^\cA_{-m,-n}$ correspondingly the \emph{singular vectors}, if $n<m+s$, in an analogy with the conformal field theory. It is evident from (\ref{Qt-commut}), (\ref{Q1-zero}) that
\eq{
{}_{mn}\langle N^{(s)}|t(X)|1\rangle_{mn}={}_{-m,-n}\langle1|t(X)|N^{(s)}\rangle_{-m,-n}=0,
\quad\text{if $s>n-m$.}
\label{Nvec-zero}
}
These identities mean that there are nonzero elements $N^{(s)}_{mn}\in\cA$ that produce the vanishing form factors:
\eq{
{}_{mn}\langle N^{(s)}|={}_{mn}\langle1|\pi(N^{(s)}_{mn}),
\qquad
|N^{(s)}\rangle_{-m,-n}=\bar\pi(N^{(s)}_{mn})|1\rangle_{-m,-n}.
\label{Nmn-def}
}
In what follows we omit the subscripts $mn$ at $N^{(s)}$, if their values will be clear from the context.

The construction of the singular vectors proposed above surprisingly resembles the construction of null vectors in~\cite{Babelon:1996sk} for form factors in the sine-Gordon models at reflectionless points. The operator $\hat{\mathcal C}$, which is the main building block of the null vectors in~\cite{Babelon:1996sk}, is expressed in terms of the fermions in just the same way as our operator $W$ in terms of to the screening modes~$S_k$. We expect that our algebraic approach can be extended to cover the reflectionless points and will give a free field realization of the construction of~\cite{Babelon:1996sk}.

It is easy to generalize the identities~(\ref{Nvec-zero}). Let
\eq{
\Aligned{
\cN^{(s)}_{mn}
&=\cF_{m,n-2s}Q^{(s)}\subseteq\cF_{mn},
\quad
&\cN^{\cA(s)}_{mn}
&=\cF^\cA_{m,n-2s}Q^{(s)}\subseteq\cF^\cA_{mn}\cap\cN^{(s)}_{mn},
\\
\bcN^{(s)}_{mn}
&=Q^{(s)}\bcF_{m,n-2s}\subseteq\bcF_{mn},
\quad
&\bcN^{\cA(s)}_{-m,-n}
&=Q^{(s)}\bcF^\cA_{-m,-n+2s}\subseteq\bcF^\cA_{-m,-n}\cap\bcN^{(s)}_{-m,-n}.
}\label{cN-def}
}
In other words, $\cN^{(s)}_{mn}$ consists of the vectors $\langle v|Q^{(s)}_{mn}$ with $\langle v|\in\cF_{m,n-2s}$, while $\cN^{\cA(s)}_{mn}$ consists of the vectors ${}_{mn}\langle N^{(s)}_h|\equiv{}_{m,n-2s}\langle h|Q^{(s)}_{mn}$ with $h\in\cA$. We have
\eq{
\langle v|t(X)|1\rangle_{mn}=0
\quad\text{$\forall\ \langle v|\in\cN^{(s)}_{mn}$, if $s>n-m$.}
\label{cN-zero}
}

It is tempting to replace the signs $\subseteq$ in (\ref{cN-def}) by the equality signs. It is not always possible. The answer depends on some properties of the screening operators. We have two conjectures based on the calculations by means of the {\sl Mathematica\/}$^\text{\textregistered}$ symbolic algebra package for the lowest levels. We formulate both conjectures for the right action on the spaces $\cF_{mn}$ and~$\cF^\cA_{mn}$. For the left action these spaces must be substituted by $\bcF_{-m,-n}$,~$\bcF^\cA_{-m,-n}$.

\begin{conjecture}
\label{conj:W-action}
For generic values of $r$ the right action of the operator $W$ from $\cF_{mn}$ to $\cF_{m,n+4}$ and from $\cF^\cA_{mn}$ to $\cF^\cA_{m,n+4}$ is surjective, if $m-n\le-2$, and indegenerate, if $m-n\ge-2$. In other words the operator $W$ is of maximal rank on each level subspace.
\end{conjecture}

The situation with the $\Sigma$ operator is more complicated. Since $\Sigma^2=0$, the operator $\Sigma$ can be considered as a differential of the complexes
$$
\Aligned{
&\cdots\stackrel\Sigma\longrightarrow(\cF_{m,n-2})_{l+m-n+1}\stackrel\Sigma\longrightarrow(\cF_{mn})_l
\stackrel\Sigma\longrightarrow(\cF_{m,n+2})_{l-m+n+1}\stackrel\Sigma\longrightarrow(\cF_{m,n+4})_{l-2m+2n+4}
\stackrel\Sigma\longrightarrow\cdots,
\\
&\cdots\stackrel\Sigma\longrightarrow(\cF^\cA_{m,n-2})_{l+m-n+1}\stackrel\Sigma\longrightarrow(\cF^\cA_{mn})_l
\stackrel\Sigma\longrightarrow(\cF^\cA_{m,n+2})_{l-m+n+1}\stackrel\Sigma\longrightarrow(\cF^\cA_{m,n+4})_{l-2m+2n+4}
\stackrel\Sigma\longrightarrow\cdots
}
$$
for its right action.

\begin{conjecture}
\label{conj:Sigma-action}
Consider the cohomology at the nod $(\cF_{mn})_l$ {\rm(}or $(\cF^\cA_{mn})_l${\rm)}. For generic values of $r$ for the first complex it is zero, if $m-n\ne0$. The cohomology at $n=m$ is nonzero at each even level\/~$l$. For the complex of $\cA$-subspaces the cohomology is zero, if $m-n\in2\Z+1$ or $m-n\in2\Z_{\ge0}$. For $m-n\in2\Z_{<0}$ the cohomology is one-dimensional at each even level\/~$l$.
\end{conjecture}

If Conjecture~\ref{conj:W-action} is true, for a generic value of $r$ and $\langle v|\in\cF_{m,n-2s}$ such that $s\ge n-m$ and $n\in2\Z$ the condition $\langle v|Q^{(s)}D_k=0$ implies $\langle v|D_k=0$ and, hence, $\cN^{\cA(s)}_{mn}=\cF^\cA_{mn}\cap\cN^{(s)}_{mn}$ for $s>n-m$ and $n\in2\Z$.

Since $Q^{(s+2)}=WQ^{(s)}$ we have $\cN^{(s+2)}_{mn}\subseteq\cN^{(s)}_{mn}$. Let $s_{mn}$ be the least value of $s$ for given $m,n$, for which the subspace $\cN^{(s)}_{mn}$ is a proper subspace of~$\cF_{mn}$:
\eq{
s_{mn}=\Cases{2,&\text{if $n\le m$, $n\in2\Z$,}\\
1,&\text{if $n\le m$, $n\in2\Z+1$,}\\
n-m+2,&\text{if $n>m$, $m\in2\Z$,}\\
n-m+1,&\text{if $n>m$, $m\in2\Z+1$.}}
\label{smn-def}
}
Then the result can be presented by the diagram:
\eq{
\begin{matrix}
\cF_{mn}&\supset&\cN^{(s_{mn})}_{mn}&\supset&\cN^{(s_{mn}+2)}_{mn}
&\supset&\cN^{(s_{mn}+4)}_{mn}&\supset&\cdots\\
\cup&&\cup&&\cup&&\cup\\
\cF^\cA_{mn}&\supset&\cN^{\cA(s_{mn})}_{mn}&\supset&\cN^{\cA(s_{mn}+2)}_{mn}&
\supset&\cN^{\cA(s_{mn}+4)}_{mn}&\supset&\cdots
\end{matrix}
\label{cN-supsets}
}
This diagram (as well as all the diagrams below) is valid for `barred' spaces after the substitution $\cF_{mn}\to\bcF_{-m,-n}$, $\cN^{(s)}_{mn}\to\bcN^{(s)}_{-m,-n}$ etc. Note that each space $\cN^{(s)}_{mn}$ (but not $\cN^{\cA(s)}_{mn}$) is a representation of the $\SVir_{q,-q}$ algebra, so that the filtration~(\ref{cN-supsets}) is a filtration of representations (more precisely, of right modules). The factors $\cF_{mn}/\cN^{(s_{mn})}_{mn}$, $\cN^{(s_{mn}+2k)}_{mn}/\cN^{(s_{mn}+2k+2)}$ are representations as well.

Now let us study the structure of the representations~$\bcF_{mn}$. It is dual to that of~$\cF_{mn}$. From (\ref{Q-gen-action}) we see that the operator $Q^{(s)}$ annihilates every vector $|w\rangle\in(\bcF_{mn})_l$ of level $l<s(m-n+s)$ and, hence, any matrix element $\langle v|t(X)|w\rangle$ for such $|w\rangle$ and $\langle v|\in\cN^{(s)}_{mn}$ vanishes. But it is not so, if $l\ge s(m-n+s)$. Consider the kernels of the operators $Q^{(s)}$:
\eq{
\bcL^{(s)}_{mn}
=\Ker_{\bcF_{mn}}Q^{(s)},
\quad
\bcL^{\cA(s)}_{mn}
=\Ker_{\bcF^\cA_{mn}}Q^{(s)},
\label{bcL-def}
}
The matrix elements $\langle v|t(X)|w\rangle$ vanish for all $\langle v|\in\cN^{(s)}_{mn}$, $|w\rangle\in\cL^{(s')}_{mn}$, if $s\ge s'$. Similarly to (\ref{cN-supsets}) we obtain the filtrations:
\eq{
\begin{matrix}
\bcL^{(s_{mn})}_{mn}&\subset&\bcL^{(s_{mn}+2)}_{mn}&\subset&\bcL^{(s_{mn}+4)}_{mn}&\subset&\cdots&
\subset&\bcF_{mn}\\
\cup&&\cup&&\cup&&&&\cup\\
\bcL^{\cA(s_{mn})}_{mn}&\subset&\bcL^{\cA(s_{mn}+2)}_{mn}&\subset&\bcL^{\cA(s_{mn}+4)}_{mn}&
\subset&\cdots&\subset&\bcF^\cA_{mn}
\end{matrix}
\label{bcL-subsets}
}
The spaces $\bcL^{(s)}_{mn}$ are again representations (left modules) of the $\SVir_{q,-q}$ algebra. The quotients $\bcL^{(s+2)}_{mn}/\allowbreak\bcL^{(s)}_{mn}$ are again modules. Nevertheless, we will be interested in other quotient spaces:
\eq{
\bcM^{(s)}_{mn}
=\bcF_{mn}/\bcL^{(s)}_{mn},
\quad
\bcM^{\cA(s)}_{mn}
=\bcF^\cA_{mn}/\bcL^{\cA(s)}_{mn},
\label{bcM-def}
}
The spaces $\bcM^{(s)}_{mn}$ are quotient modules and, hence, (reducible) representations of the $\SVir_{q,-q}$ algebra. The lowest level vector in $\bcM^{(s)}_{mn}$ lies at the level $s(m-n+s)$. We will call it a \emph{cosingular vector}. Below we will see that the cosingular vector belongs to the subspace $\bcM^{\cA(s)}_{mn}$ subject to $mn\in2\Z$, providing important identities for form factors. If both $m$ and $n$ are odd, this lowest level vector is not an $\cA$-vector.

The definition (\ref{bcM-def}) together with the filtration (\ref{bcL-subsets}) leads to the conclusion that $\bcM^{(s+2)}_{mn}=\bcM^{(s)}_{mn}/\bcL^{(s+2)}_{mn}$, $\bcM^{\cA(s+2)}_{mn}=\bcM^{\cA(s)}_{mn}/\bcL^{\cA(s+2)}_{mn}$. Hence, we have a sequence
\eq{
\begin{matrix}
\bcF_{mn}&\to&\bcM^{(s_{mn})}_{mn}&\to&\bcM^{(s_{mn}+2)}_{mn}
&\to&\bcM^{(s_{mn}+4)}_{mn}&\to&\cdots\\
\cup&&\cup&&\cup&&\cup\\
\bcF^\cA_{mn}&\to&\bcM^{\cA(s_{mn})}_{mn}&\to&\bcM^{\cA(s_{mn}+2)}_{mn}
&\to&\bcM^{\cA(s_{mn}+4)}_{mn}&\to&\cdots
\end{matrix}
\label{cM-factor}
}
where the arrows mean the quotient maps.

Now consider any matrix element $\langle v|t(X)|v'\rangle$ with $v\in\cN^{(s)}_{mn}$. If the vector $|v'\rangle$ represents an element $|w'\rangle\in\bcM^{(s')}_{mn}$ with $s'\le s$, the matrix element is representative independent and we may write $\langle v|t(X)|w'\rangle$ as well. In particular, the matrix element vanishes, if $|w'\rangle=0$. In the case $s'=s$ there is one more interesting feature. The condition that this matrix element provides a form factor is weaker than that in the general case. Namely, the vector $\langle v|$ must be an $\cA$-vector, but for the vector $|w'\rangle$ it is sufficient that it were an $\cA$-vector in the space~$\bcM^{(s)}_{mn}$. In other words, in $\cF_{mn}$ the vector $|v'\rangle$ satisfies the condition $Q^{(s)}s_k|v'\rangle=0$ or, equivalently, $Q^{(s)}D_k|v'\rangle=0$ for $k>0$. Such a `weakly $\cA$-' (but not an $\cA$-) vector may exist, if the kernel of the left action of $Q^{(s)}$ is nonzero. For even $n$ it is possible, if $s>n-m$. For odd $n$ such vectors may exist for arbitrary values of~$m,s$. An example of a weak $\cA$-vector, $(d^-_{-1}+d^+_{-1})|1\rangle_{11}$, appears below in the proof of the equation of motion in the form factor form~(\ref{eqmotion-ff}).

Due to the bijection (\ref{cFbcF-isomorph}) of the spaces $\cF_{mn}$ and $\bcF_{-m,-n}$, both structures (\ref{cN-supsets}) and (\ref{cM-factor}) exist for each module. We may introduce the spaces
\Align{
\bcN^{(s)}_{mn}
&=Q^{(s)}\bcF_{mn}\stackrel\top\longleftrightarrow\cN^{(s)}_{-m,-n},
\label{bcN-def}
\\
\cL^{(s)}_{mn}
&=\Ker_{\cF_{mn}}Q^{(s)}\stackrel\top\longleftrightarrow\bcL^{(s)}_{-m,-n},
\label{cL-def}
\\
\cM^{(s)}_{mn}
&=\cF_{mn}/\cL^{(s)}_{mn}\stackrel\top\longleftrightarrow\bcM^{(s)}_{-m,-n},
\label{cM-def}
}
where the kernel in (\ref{cL-def}) is the kernel of the \emph{left} action on the space~$\cF_{mn}$. If Conjecture~\ref{conj:W-action} is true, it means that for even values of $n$ the space $\cF_{mn}$ splits into a sum of the presumably irreducible modules $\cN^{(s)}_{mn}/\cN^{(s+2)}_{mn}\cong\cL^{(m-n+s-2)}_{mn}/\cL^{(m-n+s)}_{mn}$. For odd values of $n$ the structure of the Fock spaces is more complicated, presumably the same as the structure of those in the free field realization of the Virasoro algebra under the Felder complex~\cite{Felder:1988zp}.

\section{Integral representation for the \texorpdfstring{$t=-q$}{t=-q} Macdonald polynomials}
\label{sec:Macdonald}

In~\cite{Shiraishi:1995rp} it was shown that the singular vectors in the free field representations of the deformed Virasoro algebra $\Vir_{q,t}$ are given by the Macdonald polynomials with rectangular partitions.%
\footnote{Note that the integral representation for singular vectors given in~\cite{Shiraishi:1995rp} is incorrect. The correct one should be given in terms of Lukyanov's screening operator~\cite{Lukyanov:1996qs,Jimbo:1996vu}, which commutes with the deformed Virasoro algebra.}
Here we prove this statement for the algebra~$\SVir_{q,-q}$. Thus we obtain an integral representation for the Macdonald polynomials with rectangular partitions for~$t=-q$. The derivation follows the guidelines of~\cite{Shiraishi:1995rp}.

The Macdonald difference operator acting in the space of symmetric polynomials of $N$ variables $x_1,\ldots,x_N$ is defined as
\eq{
\hat H=\sum^N_{i=1}\left(\prod^N_{j\ne i}{tx_i-x_j\over x_i-x_j}\right)T_{q,i},
\label{hatH-def}
}
where $T_{q,i}$ is the shift operator: $T_{q,i}f(x_1,\ldots,x_i,\ldots x_N)=f(x_1,\ldots,qx_i,\ldots,x_N)$. Let $\lambda=(\lambda_1,\lambda_2,\ldots)$ be a partition of an integer $L=|\lambda|$ such that $\lambda_1\ge\lambda_2\ge\cdots\ge0$. The Macdonald polynomial $P_\lambda(X)$ associated to a partition $\lambda$ is the eigenfunction of this operator:
\eq{
\hat HP_\lambda=\ve_\lambda P_\lambda,
\qquad
\ve_\lambda=\sum^N_{i=1}t^{N-i}q^{\lambda_i},
\quad\text{if $N\ge L$.}
\label{Plambda-def}
}
This defines the Macdonald polynomial uniquely up to a normalization factor.

Now return to the case $t=-q$. Any symmetric polynomial $P(X)$ can be expressed in terms of the Newton polynomials $p_k(X)=\sum^N_{i=1}x_i^k$. Consider the isomorphism map $[\cdot]$ from the ring of symmetric polynomials to the commutative algebra~$\cA$:
\eq{
[p_k]=c_{-k}(1-q^{-k}).
\label{pk-ck}
}
In this section we prove

\begin{theorem}

The vectors generated by the screening operators from the Fock vacuum vectors are proportional to the Macdonald polynomials:
\eq{
{}_{m,n-2s}\langle1|Q^{(s)}={}_{mn}\langle[P_{s\times(m-n+s)}]|C_{s,m-n+s},
\qquad
Q^{(s)}|1\rangle_{\overline{m,n-2s}}
=C_{s,m-n+s}|[P_{s\times(m-n+s)}]\rangle_{\overline{mn}}
\label{Qs-Plambda}
}
with some numeric coefficients $C_{ss'}$, if $s>n-m$, $s-n\in2\Z$. Here $s\times s'$ denotes a rectangular partition such that $(s\times s')_i=s'$ $(1\le i\le s)$, $(s\times s')_i=0$ $(i>s)$.

\end{theorem}

Though the proof of the theorem basically repeats the proof of a similar theorem in~\cite{Shiraishi:1995rp}, there are some peculiar features in our case.

By using the map (\ref{pk-ck}) define the operators $H$ and $\bar H$ as follows:
\eq{
\langle[P]|H=\langle[\hat HP]|,
\qquad
\bar H|[P]\rangle=|[\hat HP]\rangle.
\label{H-barH-def}
}
To write down these operators explicitly in terms of the free bosons, we have to make some arrangements. Introduce the exponents
\eq{
\Psi(z)=\e^{-\i\pi\hat a}\exp\left(-\sum_{k>0}{d^+_kz^{-k}\over k}\right)=\sum_{k\ge0}\Psi_kz^{-k},
\qquad
\bar\Psi(z)=\e^{\i\pi\hat a}\exp\left(-\sum_{k<0}{d^-_kz^{-k}\over k}\right)=\sum_{k\le0}\bar\Psi_kz^{-k}.
\label{Psi-barPsi-def}
}
Note that the operator $\Psi(z)$ only contains the positive modes ($d^\pm_k$ with $k>0$), while $\bar\Psi(z)$ the negative ones. Hence
\eq{
[\pi(c_{-k}),\Psi(z)]=[\bar\pi(c_{-k}),\bar\Psi(z)]=0.
\label{ck-Psi-commut}
}
We will use the notation
\eq{
(t\Psi)_k=\oint{dz\over2\pi\i}\,z^{k-1}t(z)\Psi(z)=\sum_{l\ge0}t_{k-l}\Psi_l,
\qquad
(\bar\Psi t)_k=\oint{dz\over2\pi\i}\,z^{k-1}\bar\Psi(z)t(z)=\sum_{l\ge0}\bar\Psi_{-l}t_{l+k},
\label{tPsi-def}
}
and, similarly, $(\lambda_\pm\Psi)_k$,~$(\bar\Psi\lambda_\pm)_k$. With this notation the operators $H$, $\bar H$ look like
\eq{
H=-{(-q)^N\over q+1}((t\Psi)_0-\e^{-2\pi\i\hat a}-(-q)^{-N}),
\qquad
\bar H=-{(-q)^N\over q+1}((\bar\Psi t)_0-\e^{2\pi\i\hat a}-(-q)^{-N}).
\label{H-barH-explicit}
}

To prove identities (\ref{H-barH-explicit}) we have to consider the operator $\hat H$ as the first member $\hat H=\hat H_0$ of an infinite family of operators:
$$
\hat H_k=\sum^N_{i=1}\left(\prod^N_{\substack{j\ne i}}{tx_i-x_j\over x_i-x_j}\right)x_i^kT_{q,i},
\qquad
k\ge0.
$$
These operators satisfy the relations
\subeq{\label{hatHk-commut}
\Gather{
\hat H_k1=\sum^N_{i=1}x_i^k\prod^N_{\substack{j\ne i}}{tx_i-x_j\over x_i-x_j},
\label{hatHk1}
\\
[\hat H_k,p_l(X)]=(q^l-1)\hat H_{k+l}.
\label{hatHkpl-commut}
}}
Define the operators $H_k$ in the Heisenberg algebra according to the same rule: $\langle[P]|H_k=\langle[\hat H_kP]|$. In terms of these operators the above relations are rewritten as follows:
\subeq{\label{Hk-commut}
\Gather{
\langle1|H_k=\left\langle\left[\sum^N_{i=1}x_i^k\prod^N_{\substack{j\ne i}}{tx_i-x_j\over x_i-x_j}\right]\right|,
\label{Hk1}
\\
[\pi(c_l),H_k]=q^lH_{k+l}+(\text{$\cA$-null}),
\label{Hkcl-commut}
}}
where `$\cA$-null' mean the terms that vanish on the $\cA$-subspace. In fact, these relations define the operators $H_k$ on the $\cA$-subspace uniquely. Now we conjecture that the operators $H_k$ have the form
\eq{
H_k=-{(-q)^N\over q+1}\left(q^k(t\Psi)_k-\delta_{k,0}(\e^{-2\pi\i\hat a}+(-q)^{-N})\right),
\label{Hk-def}
}
and prove them to satisfy relations~(\ref{Hk-commut}).

By using (\ref{Apm-def}), (\ref{pk-ck}), (\ref{tPsi-def}) we immediately have
\Multline{
{}_{mn}\langle1|(t\Psi)_k={}_{mn}\langle1|\oint{dz\over2\pi\i}\,z^{k-1}
\left(\e^{-2\pi\i a_{mn}}+\exp\sum_{k>0}{A^+_k\pi(c_{-k})z^{-k}\over k}\right)
\\
=\oint{dz\over2\pi\i}\,z^{k-1}\>\prescript{}{mn\!}{\left\langle\left[
\delta_{k,0}\e^{-2\pi\i a_{mn}}+\oint{dz\over2\pi\i}z^{k-1}\prod^N_{i=1}{z+x_i\over z-qx_i}\right]\right|}.
\?}
The integration contour of the integral in the r.~h.~s.\ is a circle around infinity, i.~e.\ a large circle that encloses all poles at the points $qx_i$ and (for $k=0$) at zero. By summing up the residues, we reduce the integral to the sum in the r.h.s.\ of~(\ref{Hk1}).

It is a little more involved to obtain the commutation relation~(\ref{Hkcl-commut}). We will use the identities
\eq{
[\pi(c_{-l}),(\lambda_\pm\Psi)_k]=(\mp)^{l+1}(\lambda_\pm\Psi)_{k+l},
\label{pi-lambdaPsi-commut}
}
which immediately follow from~(\ref{pibpi-props}).

First of all, notice that
\eq{
{}_{mn}\langle h|(\lambda_+\Psi)_k={}_{mn}\langle h|\delta_{k0}\e^{-2\pi\i a_{mn}},
\qquad
k\ge0,
\qquad
h\in\cA.
\label{hvec-lambda+Psi}
}
Indeed, for the vacuum vector we have
\Multline{
{}_{mn}\langle1|(\lambda_+\Psi)_k=
{}_{mn}\langle1|\oint{dz\over2\pi\i}z^{k-1}\e^{-2\pi\i a_{mn}}\exp\sum_{l<0}{d^+_lz^{-l}\over l}
={}_{mn}\langle1|\oint{dz\over2\pi\i}z^{k-1}\e^{-\i\pi a_{mn}}
\\*
={}_{mn}\langle1|\delta_{k0}\e^{-\i\pi a_{mn}},
\?}
since ${}_{mn}\langle1|d^+_l=0$ for $l<0$. On the other hand, form (\ref{pi-lambdaPsi-commut}) we have
$$
\pi(c_{-l})(\lambda_+\Psi)_k=(\lambda_+\Psi)_k\pi(c_{-l})+(-)^{k+1}(\lambda_+\Psi)_{k+l}.
$$
Hence the subscript of $(\lambda_+\Psi)_k$ may only increase while we move the operator to the left. The only term that admits a nonzero contribution is the first one. In other words, the commutators vanish on the $\cA$-subspace. It proves~(\ref{hvec-lambda+Psi}).

Now, it is clear that
$$
H_k=-{(-q)^Nq^k\over q+1}\e^{\i\pi\hat a}(\lambda_-\Psi)_k+(\text{$\cA$-null}),
\quad\text{if $k>0$.}
$$
By using (\ref{pi-lambdaPsi-commut}) we immediately obtain that the operators $H_k$ of the form (\ref{Hk-def}) satisfy~(\ref{Hkcl-commut}).

Now let us show that the vectors ${}_{mn}\langle1|Q^{(s)}$ are eigenvectors of the operator $H$ with appropriate eigenvalues for $s-n\in2\Z$. It is easy to check that
\eq{
\ve_{s\times s'}=-{(-q)^N\over q+1}\left((-q)^{-s}+q^{s'}-q^{s'}(-q)^{-s}-(-q)^{-N}\right).
\label{ve-rect}
}
For $s-n\in2\Z$ we can use the commutativity of $Q^{(s)}$ and $t(z)$ to calculate the action of~$H$:
$$
{}_{m,n-2s}\langle1|Q^{(s)}H
=-{(-q)^N\over q+1}{\>}_{m,n-2s}\langle1|\left(\sum_{k\ge0}t_{-k}Q^{(s)}\Psi_k
-Q^{(s)}\left(\e^{-2\pi a_{mn}}+(-q)^{-N}\right)\right).
$$
Since ${}_{mn}\langle1|t_{-k}=0$ for $k>0$ and ${}_{mn}\langle1|t_0=2\cos\pi a_{mn}$, the only term of the sum over $k$ remains and we have
$$
{}_{m,n-2s}\langle1|Q^{(s)}H
=-{(-q)^N\over q+1}{\>}_{m,n-2s}\langle1|Q^{(s)}\left(2\e^{-\i\pi a_{mn}}\cos\pi a_{m,n-2s}
-\e^{-2\pi\i a_{mn}}-(-q)^N\right).
$$
For $s-n\in2\Z$ the eigenvalue in the r.~h.~s.\ coincides with~(\ref{ve-rect}) with $s'=m-n+s$.

\textbf{Remark 1.} The proof for values of $s$ different from $n$ is only applicable in the case $t=-q$. We have mentioned that the possibility of $s\ne n$ is related to the reflection equations and quasiperiodicity. The quasiperiodicity property remains valid for an arbitrary $t\ne-q$, while the reflection property seems to be specific to the case $t=-q$.

\textbf{Remark 2.} By using the identification (\ref{pk-ck}) we obtain the following integral expression for the Macdonald polynomials with the rectangular diagrams:
\Multline{
P_{s\times s'}(X)=C_{ss'}^{-1}\prod^{s-1}_{i=1}\oint{dw_i\over2\pi\i w_i}\,
w_i^{-(s-i)(s'+i)}\prod_{1\le i\le j<s}\left(1-\prod^j_{t=i}w_t^2\right)
\prod^{\lfloor s/2\rfloor}_{i=1}F^s(w_{2i-1})\times
\\*
\times\Res_{z=0}z^{-ss'-1}\exp\sum^\infty_{k=1}{q^{k/2}-(-)^kq^{-k/2}\over k(q^{k/2}-q^{-k/2})}
q^{k/2}\left(1+\sum^{s-1}_{i=1}\prod^i_{j=1}w_j^k\right)z^kp_k(X),
\label{Pss'-int}
}
where the contours are small circles enclosing zero. Notice, that this formula is valid for every value of~$q$ including the unit circle ($|q|=1$) except for the point $q=1$. Indeed, the last residue simply singles out the terms in the exponent that are of the $ss'$ total degree in~$X$. There is a finite number of such terms, and it is easy to write them explicitly. Also we note that this is a particular feature of the case $t=-q$, where the products of the screening currents, being written in terms of the normal products, do not contain the $q$-gamma functions. In the general case the $q$-gamma functions, being undefined on the unite circle, spoil the integral representation for~$|q|=1$.

\section{Some physical applications}
\label{sec:phys}

Here we will use the screening operators to derive some identities between form factors in the sinh-Gordon model. First, we derive the resonance identities of some descendant operators over the fields $V_{mn}(x)$ with even~$mn$. Second, we give a simple derivation of the equation of motion. And, third, we derive the conservation laws. This provides physical applications to our algebraic construction.

\subsection{Resonance identities}
\label{subsec:phys:resonance}

It is clear from our construction that the matrix elements of the form $\langle v|t(X)|w\rangle$ may occur nonzero, if $\langle v|\in\cN^{(s)}_{mn}$ and $|w\rangle$ represents a nonzero vector in~$\bcM^{(s)}_{mn}$. Let us choose the singular vector ${}_{mn}\langle N^{(s)}|$ as~$\langle v|$. We know that it is a level $s(m-n+s)$ $\cA$-vector. There must be a level $s(m-n+s)$ vector $|w\rangle$ to produce a nonzero matrix element. The question to be posed is whether it is an $\cA$-vector as well. We will see just now that it is an $\cA$-vector, if and only if the product $mn$ is even.

Consider first the simplest case $s=1$, $n\in2\Z+1$. The singular vector ${}_{mn}\langle N^{(1)}|$ lies on the level $L=m-n+1$. Correspondingly, we are looking for a level $L$ vector in $\bcF_{mn}$, which will represent the cosingular vector, i.~e.\ the highest weight vector of~$\cM^{(1)}_{mn}$. Let us try the vector $|c_{-L}\rangle_{mn}$. Since $[S(z),d^-_{-k}-d^+_{-k}]=-{A^+_k+A^-_k\over q^{k/2}-q^{-k/2}}z^{-k}S(z)$, we have
\eq{
Q^{(1)}|c_{-L}\rangle_{mn}=\kappa_L|1\rangle_{m,n-2},
\qquad
\kappa_k=\Cases{0,&k\in2\Z+1,\\-2/(q^{k/2}-q^{-k/2}),&k\in2\Z.}
\label{Q1cL-id}
}
For even values of~$L$, which correspond to even~$m$, the vector $|c_{-L}\rangle_{mn}$ represents the cosingular vector, which belongs to~$\bcM^{\cA(1)}_{mn}$.

For odd values of $L$ the r.~h.~s.\ vanishes, which means that the form factors $f^{N^{(1)}\bar c_{-L}}_{mn}$ vanish. It is easy to check that for odd $L$ any $\cA$-vector $|c_{-\vec k}\rangle_{mn}=|c_{-k_1}\cdots c_{-k_\nu}\rangle_{mn}$, $\sum k_i=L$, is annihilated by~$\Sigma$. Indeed, at least one of the integers $k_i$, say, $k_\nu$ is odd in this case. Hence, the most singular term in the commutator $[S[z],\bar\pi(c_{-\vec k})]$ is proportional to $z^{-(k_1+\cdots+k_{\nu-1})}S(z)$. But $k_1+\dots+k_{\nu-1}<k_1+\dots+k_{\nu-1}+k_\nu=L$ and, hence, the exponent of $z$ in the integrand is greater than~$-1$. Thus the integral vanishes. It means that the cosingular vector does not belong to~$\cM^{\cA(1)}_{mn}$. For example, it can be represented by the vector $(d^-_{-L}+d^+_{-L})|1\rangle_{mn}$. Indeed,
$$
Q^{(1)}(d^-_{-L}+d^+_{-L})|1\rangle_{mn}=-{2A^+_L\over q^{L/2}-q^{-L/2}}|1\rangle_{mn}.
$$
Surely, the vector $(d^-_{-L}+d^+_{-L})|1\rangle_{mn}$ is an example of a `weak' $\cA$-vector. The case $L=1$ will be studied in more detail in the next subsection. Note that similar argument can be applied to the vectors of odd level $L=s(m-n+s)$ for arbitrary (odd) values of~$s$.

Return to the case of even values of~$L$. From the identity
$$
{}_{mn}\langle N^{(1)}|t(X)|c_{-L}\rangle_{mn}
={}_{m,n-2}\langle1|\Sigma t(X)|c_{-L}\rangle_{mn}
={}_{m,n-2}\langle1|t(X)\Sigma|c_{-L}\rangle_{mn}
=\kappa_L\times{}_{m,n-2}\langle1|t(X)|1\rangle_{m,n-2}
$$
we immediately obtain
\eq{
F^{h^{(1)}_{mn}}_{mn}(\theta_1,\ldots,\theta_N)
=\kappa_LF_{m,n-2}(\theta_1,\ldots,\theta_N),
\qquad
h^{(1)}_{mn}=N^{(1)}_{mn}\bar c_{-L}.
\label{f(1)-id}
}
To clarify the meaning of the identity, let us stir the parameter $a$ near the value~$a_{mn}$. Then we obtain
\eq{
F^{h^{(1)}_{mn}}_a(\theta_1,\ldots,\theta_N)
=\kappa_LF_{a+r-1}(\theta_1,\ldots,\theta_N)+O(a-a_{mn})
\label{f(1)-stir}
}
as $a\to a_{mn}$. The operator $V^{h^{(1)}_{mn}}_a$ defined by the form factors in the l.~h.~s.\ are normalized unnaturally from the point of view of the Lagrangian theory. Indeed, it is proportional to a Lagrangian operator of the form
$$
\cO^{(1)}_a(x)=\left((\d\varphi)^L(\bd\varphi)^L+\cdots\right)\e^{\alpha\varphi},
$$
where the dots stand for the terms that contain higher derivatives of~$\varphi$. But the vacuum expectation value of the operator $\cO^{(1)}_a$ is known to have a pole at $a=a_{mn}$~\cite{Fateev:1998xb,Jimbo:2009ja}. The origin of this pole is related to the operator resonances~\cite{Zamolodchikov:1990bk,Fateev:1998xb,Lashkevich:2011ne}. If $F^{(1)}_a(\theta_1,\ldots,\theta_N)$ are the corresponding form factors, the identity (\ref{f(1)-stir}) takes the usual form of a resonance identity
\eq{
F^{(1)}_a(\theta_1,\ldots,\theta_N)
={\const\over a-a_{mn}}F_{a+r-1}(\theta_1,\ldots,\theta_N)+O(1)
\quad\text{as $a\to a_{mn}$.}
\label{f(1)-res}
}
Evidently, $V^{h^{(1)}_{mn}}_a(x)\propto(a-a_{mn})\cO^{(1)}_a(x)$.

Now let us generalize the identities~(\ref{f(1)-id}) to an arbitrary value of~$s$ subject to even~$L$.

To do this, we will 
need the particular property of the operator $\overline Q^{(n-m)}$ (recall that $\overline Q^{(s)}$ is the same as $Q^{(s)}$ but may act on the space $\bcF_{mn}$ with odd difference $s-n$ as well) on the space $\bcF_{mn}$: it does not change the level of a vector. In particular,
\eq{
\overline Q^{(n-m)}|1\rangle_{mn}=\kappa^{(n-m)}_m|1\rangle_{m,2m-n},
\label{Q(n-m)-id}
}
where
$$
\kappa^{(s)}_m=\prod^{s-1}_{i=1}\oint{dw_i\over2\pi\i w_i}\,w_i^{-i(s-i)}
\prod_{1\le i\le j<s}\left(1-\prod^j_{t=i}w_t^2\right)\prod^{\lfloor s/2\rfloor}_{i=1}F^{m+s}(w_{2i-1}).
$$
It can be checked by means of {\it Mathematica$^\text{\textregistered}$} (up to $s=9$) that
\eq{
\kappa^{(2k)}_m=k!\prod^k_{i=1}F^m_{2i-1},
\qquad
\kappa^{(2k+1)}_m=(-)^kk!\prod^k_{i=1}F^{m+1}_{2i}.
\label{kappasm-explicit}
}

Let $\vec k=(k_1,\ldots,k_\nu)$ and denote $S_{-\vec k}=S_{-k_1}\cdots S_{-k_\nu}$ (the order is essential). We are interested in the vectors
\eq{
S_{-\vec k}\overline Q^{(t)}|1\rangle_{m,m+s+t}\in\bcF^\cA_{mn},
\qquad
\nu={m-n+s-t\over2}\ge0,
\qquad
\sum^\nu_{i=1}k_i=s(m-n+s-t).
\label{Skvec-even-def}
}
Notice that, since $s-n$ is even, the difference $t-m$ is even as well. Therefore, the vector $\overline Q^{(t)}|1\rangle_{m,m+s+t}$ is a singular vector in $\bcF_{m,m+s-t}$ subject to $m-n\in2\Z$. In the case $m-n\in2\Z+1$ the operator $\overline Q^{(t)}$ does not commute with the algebra~$\SVir_{q,-q}$ and the corresponding vector is not a singular vector.

Using the commutation relations (\ref{Sk-commut}) we immediately obtain
$$
Q^{(s)}S_{-\vec k}\overline Q^{(t)}|1\rangle_{m,m+s+t}
=(-)^{\nu s}S_{-\vec k+2s\vec1}\overline Q^{(s+t)}|1\rangle_{m,m+s+t}
=(-)^{\nu s}\kappa^{(s+t)}_mS_{-\vec k+2s\vec1}|1\rangle_{m,m-s-t},
$$
where $\vec1=(1,\ldots,1)$, if at least one of the values $s$ or $t$ is even. The vector $S_{-\vec k+2s\vec1}|1\rangle_{m,m-s}$ in the r.~h.~s.\ being a zero level vector is proportional to $|1\rangle_{mn}$:
\eq{
S_{-\vec k+2s\vec1}|1\rangle_{m,m-s}=C^{(r)}_{\vec k-2s\vec1}|1\rangle_{m,n-2s},
\qquad
C^{(\nu)}_{\vec k}=\prod^{\nu-1}_{i=1}\oint{dz_i\over2\pi\i z_i}\,
z_i^{-\sum^\nu_{t=i+1}k_i}\prod_{1\le i<j<\nu}\left(1-\prod^j_{t=i}z_t^2\right).
\label{crk-def}
}
Finally, we have
\Multline{
Q^{(s)}S_{-\vec k}\overline Q^{(t)}|1\rangle_{m,m+s+t}
=(-)^{\nu s}\kappa^{(s+t)}_mC^{(\nu)}_{\vec k-2s\vec1}|1\rangle_{m,n-2s},
\\
\nu={m-n+s-t\over2},
\qquad
\text{if $s(m-n+s)\in2\Z$.}
\label{Qs|Sk-even}
}
In the case of odd level $s(m-n+s)$ our construction give just zero (since $s$ and $t$ are odd and $Q^{(s)}Q^{(t)}=0$ in this case) in accordance with the above remark about the absence of a cosingular $\cA$-vector on an odd level.

The coefficients $C^{(\nu)}_{\vec k}$ are all equal to $0,\pm1$. They can only be nonzero for even $k_i$ satisfying the condition $0\le k_i+k_{i+1}+\cdots+k_\nu\le2(i-1)(\nu-i+1)$. For example, $c_{\vec0}=1$, which gives the identity
\eq{
Q^{(s)}S_{-2s}^\nu\overline Q^{(t)}|1\rangle_{m,m+s+t}=(-)^{\nu s}\kappa^{(s+t)}_m|1\rangle_{m,n-2s},
\label{Qs|Sk-even-spec}
}
so that $S_{-2s}^\nu\overline Q^{(t)}|1\rangle_{m,m+s+t}$ is a simple representative of the cosingular vector.

It is interesting to note two examples of the representatives of this form. The first example corresponds to $t=m-n+s$:
\eq{
\bar\pi(M_{mn}^{(s)})|1\rangle_{mn}=|M^{(s)}\rangle_{mn}
=(\kappa^{(m-n+2s)}_m)^{-1}\overline Q^{(m-n+s)}|1\rangle_{m,2m-n+2s}.
\label{Mmn-def}
}
The representative $|M^{(s)}\rangle_{mn}$ is remarkable for, if $m$ and $n$ are even, it is in the same time a singular vector in~$\bcF_{mn}$:
$$
M^{(s)}_{mn}=N^{(m-n+s)}_{-m,-n},
\quad\text{if $m,n\in2\Z$.}
$$

By using (\ref{Qs|Sk-even-spec}) we obtain
\eq{
{}_{mn}\langle N^{(s)}|t(X)|M^{(s)}\rangle_{mn}={}_{m,n-2s}\langle1|t(X)|1\rangle_{m,n-2s}
\label{J(s)-id}
}
and
\eq{
F^{N^{(s)}\overline{M^{(s)}}}_{mn}(\theta_1,\ldots,\theta_N)
=F_{m,n-2s}(\theta_1,\ldots,\theta_N).
\label{f(s)-id}
}
In the same manner as (\ref{f(1)-id}) this identity is related to an operator resonance.

The second example of representative of the cosingular vector corresponds to $t=0$ for $m-n+s\in2\Z$ and $t=1$ for $m-n+s\in2\Z+1$, $s\in2\Z$:
$$
\bar\pi(M_{mn}^{\prime(s)})|1\rangle_{mn}=|M^{\prime(s)}\rangle_{mn}
=\Cases{(\kappa^{(s)}_m)^{-1}S_{-2s}^r|1\rangle_{m,m+s},&m-n+s=2r\in2\Z,\\
(\kappa^{(s+1)}_m)^{-1}S_{-2s}^r\Sigma|1\rangle_{m,m+s+1},&m-n+s=2r+1\in2\Z+1,\quad s\in2\Z.}
$$
The `longest' representative $M^{(s)}$ corresponds to the maximal value of $t$, while the `shortest' representative $M^{\prime(s)}$ corresponds to the minimal one.

\subsection{Equation of motion}
\label{subsec:phys:eqmo}

The treatment of operator resonances for odd levels $s(m-n+s)$ are much more complicated than that for even levels~\cite{Lashkevich:2011ne}. We have no general construction for these resonances. Nevertheless, the level~1 resonance identity, known as the equation of motion, can be rather easily obtained. In~\cite{Babujian:2002fi} this equation of motion was proved directly by using explicit expression for the breather form factors. Here we give an alternative proof based on the free field representation technique. There are two advantages of the proof we propose here. First, it is very simple. Second, we expect that it be possible to generalize it to higher odd level resonances and, which may occur yet more interesting, to the kink form factors.

For the level one singular vector we have $s=1$, $n=m\in2\Z+1$. Without loss of generality we may take $m=n=1$. Indeed, $a_{1+2k,1+2k}=a_{11}+k$. Due to quasiperiodicity (\ref{Jhh'-periodicity}) the form factors for to $a_{1+2k,1+2k}$ are easily expressed in terms of those for~$a_{11}$.

First of all, we have
\eq{
\left.{d\over da}{\>}_a\langle1|t(X)|1\rangle_a\right|_{a=a_{11}=1/2}=0,
\quad\text{if $N=\#X\in2\Z$.}
\label{dfa11evenN=0}
}
This immediately follows from the reflection property (\ref{ts-reflection}) and quasiperiodicity~(\ref{Jhh'-periodicity}), and conforms to the fact that it is proportional to the form factor of the field~$\varphi(x)$.

Now let $N\in2\Z+1$. It is evident that
\eq{
0=\left.{d\over da}{\>}_a\langle c_{-1}|t(X)t(z)|1\rangle_a\right|_{a=a_{11}}
={}_{11}\langle c_{-1}|{dt(X)\over da}t(z)|1\rangle_{11}
+{}_{11}\langle c_{-1}|t(X){dt(z)\over da}|1\rangle_{11}.
\label{dfXz=0}
}
Expand the r.~h.~s.\ in the variable~$z$ near $z=0$. Evidently,
$$
t(z)|1\rangle_a=(2\cos\pi a-(\e^{\i\pi a}d^-_{-1}+\e^{-\i\pi a}d^+_{-1})z)|1\rangle_a+O(z^2),
$$
and, hence,
$$
\Gathered{
t(z)|1\rangle_{11}=-\i z(d^-_{-1}-d^+_{-1})|1\rangle_{11}+O(z^2)
=-\i zA^+_1|c_{-1}\rangle_{11}+O(z^2),
\\
{dt(z)\over da}|1\rangle_{11}=(-2\pi+\pi z(d^-_{-1}+d^+_{-1}))|1\rangle_{11}+O(z^2).
}
$$
Substituting this to the r.~h.~s.\ of (\ref{dfXz=0}) and taking the coefficient at $z^1$ we obtain
$$
0=-\i A^+_1\left.{d\over da}{\>}_a\langle c_{-1}|t(X)|c_{-1}\rangle_a\right|_{a=a_{11}}
+\pi{\>}_{11}\langle c_{-1}|t(X)(d^-_{-1}+d^+_{-1})|1\rangle_{11}.
$$
Using ${}_{11}\langle c_{-1}|=B_1^{-1}{\>}_{1,-1}\langle1|Q^{(1)}$ and $Q^{(1)}|_{\bcF_{11}}=S_0$, we obtain
\Multline{
\left.{d\over da}{\>}_a\langle c_{-1}|t(X)|c_{-1}\rangle_a\right|_{a=a_{11}}
=-{\i\pi\over A^+_1}{\>}_{11}\langle c_{-1}|t(X)(d^-_{-1}+d^+_{-1})|1\rangle_{11}
\\
=-{\i\pi\over B_1A^+_1}{\>}_{1,-1}\langle1|Q^{(1)}t(X)(d^-_{-1}+d^+_{-1})|1\rangle_{11}
=-{\i\pi\over B_1A^+_1}{\>}_{1,-1}\langle1|t(X)Q^{(1)}(d^-_{-1}+d^+_{-1})|1\rangle_{11}
\\
=-{\i\pi\over B_1A^+_1}{\>}_{1,-1}\langle1|t(X)[S_1,d^-_{-1}+d^+_{-1}]|1\rangle_{11}
={\i\pi\over\sin\pi r}{\>}_{1,-1}\langle1|t(X)S_0|1\rangle_{11}
\\
=-{\pi\over\sin\pi r}{\>}_{1,-1}\langle1|t(X)|1\rangle_{1,-1},
\?}
which proves the identity
\eq{
\left.{d\over da}{\>}_a\langle c_{-1}|t(X)|c_{-1}\rangle_a\right|_{a=a_{11}}
=-{\pi\over\sin\pi r}{\>}_{1,-1}\langle1|t(X)|1\rangle_{1,-1},
\qquad\text{if $N\in2\Z+1$.}
\label{eqmotion-ff}
}
Taking into account the periodicity, we get the identity
\eq{
V^{\prime c_{-1}\bc_{-1}}_{11}(x)=-{\pi\over2G_{13}\sin\pi r}(V_{1,-1}(x)-V_{13}(x)).
\label{eqmotion-V}
}
The l.~h.~s.\ is proportional to $\d\bd\varphi$. Restoring the coefficient, we obtain the equation of motion
\eq{
\d\bd\varphi=4\pi\mu b\sh b\varphi
\label{eqmotion-varphi}
}
(see Appendix~A.1 of~\cite{Feigin:2008hs} for details).

\subsection{Conservation laws}
\label{subsec:phys:cl}

Any integrable field theory contains an infinite set of integrals of motion. The corresponding conservation laws can be written in the form of the continuity equation. In the case of the sinh-Gordon model they can be written as
\eq{
\bd T_{2k+2}=\d\Theta_{2k},
\qquad
\d T_{-2k-2}=\bd\Theta_{-2k},
\qquad
k=0,1,2,\ldots.
\label{conservlaws}
}
The subscript means the (Lorentz) spin of the corresponding current. The first currents are proportional to components of the energy-momentum tensor:
\eq{
\Aligned{
T_2
&=-2\pi T_{zz}=-{1\over4}(\d\varphi)^2,
\\
T_{-2}
&=-2\pi T_{\bz\bz}=-{1\over4}(\bd\varphi)^2,
\\
\Theta_0
&=2\pi T_{z\bz}=-{2\pi\mu\over r}\ch b\varphi.
}\label{em-tensor}
}
In practice, it is more useful to use the currents $T^+_{2k}$, $\Theta^+_{2k}$ or $T^-_{2k}$, $\Theta^-_{2k}$, which contain extra full derivatives~\cite{Dotsenko:1984nm} and admit generating functions~\cite{Bazhanov:1994ft,Bazhanov:1998dq}. In particular,
\eq{
\Aligned{
T^\pm_2
&=-{1\over4}(\d\varphi)^2\pm{Q\over2}\d^2\varphi,
\\
T^\pm_{-2}
&=-{1\over4}(\bd\varphi)^2\pm{Q\over2}\bd^2\varphi,
\\
\Theta^\pm_0
&=-{2\pi\mu\over r}\e^{\mp b\varphi}.
}\label{em-pm-tensor}
}
The currents $T^+_{\pm2}$ generate the two chiral Virasoro algebras of the Liouville theory~(\ref{Liouv-action}), while $T^-_{\pm2}$ those of the Liouville theory, obtained from (\ref{Liouv-action}) by the substitution $\varphi\to-\varphi$.

The currents $\Theta^+_0$ and $\Theta^-_0$ are proportional to $V_{13}$ and $V_{1,-1}$ correspondingly. Generally, the currents $\Theta^+_{2k}(x)$ are descendants of $V_{13}$, while $\Theta^-_{2k}(x)$ of $V_{1,-1}$. Here we will obtain form factors of $T^+_{2k+2}$, $\Theta^+_{2k}$, $T^-_{-2k-2}$, $\Theta^-_{-2k}$ for $k=0,1,2,\ldots$.

In what follows we need the elements $\ft_{-k}\in\cA$ defined as
\eq{
{}_a\langle\ft_{-k}|={}_{a-1+r}\langle1|S_k
\quad\text{or}\quad
S_{-k}|1\rangle_{a+1-r}=(-)^k|\ft_{-k}\rangle_a.
\label{tk-states-def}
}
They are obtained by the generating function
\eq{
\sum^\infty_{k=1}\ft_{-k}z^k=\exp\sum^\infty_{k=0}B_kc_{-k}z^k,
\qquad
B_k={q^{k/2}-(-)^kq^{-k/2}\over k}.
\label{tk-states-genfunc}
}
Explicitly, we have, for example,
$$
\Gathered{
\ft_0=1,
\qquad
\ft_{-1}=B_1c_{-1},
\qquad
\ft_{-2}=B_2c_{-2}+{1\over2}B_1^2c_{-1}^2,
\\
\ft_{-3}=B_3c_{-3}+B_2B_1c_{-2}c_{-1}+{1\over6}B_1^3c_{-1}^3,
\\
\ft_{-4}=B_4c_{-4}+B_3B_1c_{-3}c_{-1}+{1\over2}B_2^2c_{-2}^2
+{1\over2}B_2B_1^2c_{-2}c_{-1}^2+{1\over4!}B_1^4c_{-1}^4.
}
$$

Since the operator $\Theta^+_0$ is an exponential field, its form factors are expressed in terms of the matrix elements ${}_{13}\langle1|t(X)|1\rangle_{13}$. The derivative $\d$ means that we have to insert $\pi(c_{-1})$ (times $\i m/2$), so that we are interesting in the matrix element
\Multline{
{}_{13}\langle c_{-1}|t(X)|1\rangle_{13}={}_{11}\langle c_{-1}|S_0t(X)|1\rangle_{13}
=B_1^{-1}{\>}_{11}\langle1|S_1S_0t(X)|1\rangle_{13}
=-B_1^{-1}{\>}_{1,-1}\langle1|S_2S_{-1}t(X)|1\rangle_{13}
\\
=-B_1^{-1}{\>}_{1,-1}\langle1|S_2Q^{(1)}t(X)|1\rangle_{13}
=-B_1^{-1}{\>}_{1,-1}\langle1|S_2t(X)Q^{(1)}|1\rangle_{13}
=-{\>}_{1,-1}\langle1|S_2t(X)|c_{-1}\rangle_{11}
\\
=-B_2\cdot{}_{11}\langle c_{-2}|t(X)|c_{-1}\rangle_{11}.
\?}
We have used the fact that ${}_{11}\langle c_{-1}^2|t(X)|c_{-1}\rangle_{11}=0$. The last expression corresponds (up to a factor) to~$\bd V^{c_{-2}}_{11}(x)$, which is identified with~$\bd T^+_2(x)$. Restoring all the coefficients (see Appendix~A.2 of~\cite{Feigin:2008hs} for details), we obtain
$$
T^+_2(x)={\i\pi m^2\over8}V^{c_{-2}}_{11}(x).
$$

Now we will show that
\eq{
T^+_{2k+2}(x)={\i\pi m^{2k+2}\over8B_2}V^{\ft_{-2k-2}}_{11}(x),
\qquad
\Theta^+_{2k}(x)=-{2\pi\mu m^{2k}\over r}V^{\ft_{-2k}}_{13}(x).
\label{TTheta-posspin}
}
The overall normalization for $k>0$ is chosen arbitrarily. It only conforms the normalization the energy-momentum tensor and the dimensions of the operators. Since the operator $S_{-1}=Q^{(1)}$ commutes with $t(X)$ on the space $\cF_{13}$, we have
\Multline{
{}_{11}\langle\ft_{-2k-2}|t(X)|c_{-1}\rangle_{11}
=B_1^{-1}{}_{1,-1}\langle1|S_{2k+2}t(X)S_{-1}|1\rangle_{13}
=B_1^{-1}{}_{1,-1}\langle1|S_{2k+2}S_{-1}t(X)|1\rangle_{13}
\\
=-B_1^{-1}{}_{1,-1}\langle1|S_1S_{2k}t(X)|1\rangle_{13}
=-{}_{11}\langle1|\pi(c_{-1})S_{2k}t(X)|1\rangle_{13}
=-{}_{11}\langle1|S_{2k}\pi(c_{-1})t(X)|1\rangle_{13}
\\
=-{}_{13}\langle c_{-1}\ft_{-2k}|t(X)|1\rangle_{13},
\label{TTheta-proof}
}
or
\eq{
V^{\ft_{-2k-2}\bc_{-1}}_{11}(x)=-G_{13}^{-1}V^{\ft_{-2k}c_{-1}}_{13}(x),
\label{TTheta-V}
}
which is equivalent to the continuity equations for the currents~(\ref{TTheta-posspin}).

For negative spins in a similar way we have
\eq{
V^{c_{-1}\bar\ft_{-2k-2}}_{11}(x)=-G_{1,-1}^{-1}V^{\overline{\ft_{-2k}c_{-1}}}_{1,-1}(x),
\label{TTheta-V-negspin}
}
and, hence,
\eq{
T^-_{-2k-2}(x)={\i\pi m^{2k+2}\over8B_2}V^{\bar\ft_{-2k-2}}_{11}(x),
\qquad
\Theta^-_{-2k}(x)=-{2\pi\mu m^{2k}\over r}V^{\bar\ft_{-2k}}_{1,-1}(x).
\label{TTheta-negspin}
}

It is not very convenient that positive and negative spin currents are of different kinds: $T^+_s$, $\Theta^+_s$ for positive values of the spin $s$, and $T^-_s$, $\Theta^-_s$ for negative ones. Surely, we would like to have explicit expressions for the currents of the same kind, e.~g.\ for $T^+_s$,~$\Theta^+_s$. Thus we have to derive the continuity equations for them for negative~$s$. To do it let us consider descendants of~$V_{-1,-1}(x)$, which, due to the reflection equation, coincides with~$V_{11}(x)$. Similarly to (\ref{TTheta-proof}), we obtain
\eq{
G_{-1,-1}^{-1}V^{\bar\ft_{-2k-2}c_{-1}}_{-1,-1}(x)=-G_{-1,-3}^{-1}V^{\overline{\ft_{-2k}c_{-1}}}_{-1,-3}(x).
\label{TThetaplus-negspin}
}
By making the reflection, we obtain
\eq{
V^{c_{-1}\overline{r_{11}(\ft_{-2k-2})}}_{11}(x)
=-G_{13}^{-1}V^{\overline{r_{13}(\ft_{-2k})c_{-1}}}_{13}(x).
\label{TThetaplus-11-negspin}
}
where $r_{mn}(h)=r_{a_{mn}}(h)$. It means that
\eq{
\Aligned{
T^+_{-2k-2}(x)
&={\i\pi m^{2k+2}\over8B_2}G_{-1,-1}^{-1}V^{\bar\ft_{-2k-2}}_{-1,-1}(x)
={\i\pi m^{2k+2}\over8B_2}V^{\overline{r_{11}(\ft_{-2k-2})}}_{11}(x),
\\
\Theta^+_{-2k}(x)
&=-{2\pi\mu m^{2k}\over r}{G_{13}\over G_{-1,-3}}V^{\bar\ft_{-2k}}_{-1,1}(x)
=-{2\pi\mu m^{2k}\over r}V^{\overline{r_{13}(\ft_{-2k})}}_{13}(x).
}\label{TTheta-plus-posspin}
}
Unfortunately, we do not know the explicit form of the reflection map (for levels higher than~3) and we cannot make use of the last equalities in each line. Moreover, it is singular as $a\to a_{11}=1/2$. Indeed, from the explicit expression (\ref{h2-reflection}) we obtain
\eq{
r_a(c_{-2})=c_{-2}-{2\i\over\pi(a-1/2)}c_{-1}^2
\quad\text{as $a\to1/2$.}
\label{refl2-singular}
}
It means that for generic $h\in\cA$ the operator $V_a^{r_{-a}(h\bc_{-2})}=V_a^{r_{-a}(h)\overline{r_a(c_{-2})}}$ is undefined at the point $a=1/2$. Nevertheless, for $h=1$ (and for any $h$ built of the odd generators $c_{-2k-1}$) it is not so. Indeed, the operator $V_a^{\bc_{-1}^2}\sim\bd^2 V_a$ vanishes as $a\to1/2$, so that
\eq{
V^{\overline{r_{11}(c_{-2})}}_{11}(x)=V^{\bc_{-2}}_{11}(x)-{2\i\over\pi}V^{\prime\bc_{-1}^2}_{11}(x),
\label{c2-refl11}
}
where the prime means the $a$-derivative.

To generalize the last expression let us consider the functions
$$
D^h_a(X)=J^h_a(X)+J^h_{-a}(X)
={}_a\langle h|t(X)|1\rangle_a+{}_{-a}\langle h|t(X)|1\rangle_{-a}.
$$
These functions evidently satisfy the equations
$$
D^h_a(X)=D^h_{-a}(X),
\qquad
D^h_a(X)=(-)^ND^h_{1-a}(X),
$$
whence we immediately obtain that $D^{\prime h}_{a_{11}}=0$, if $N\in2\Z$. Since this property is analogous to~(\ref{dfa11evenN=0}), we may apply to this function the arguments we used earlier to derive the equation of motion. Let us define linear operators $\tau_{-k}$ on $\cA$ as
\eq{
{}_a\langle\tau_{-k}(h)|={}_{a-(1-r)}\langle h|S_k.
\label{tauk-def}
}
Evidently $\ft_{-k}=\tau_{-k}(1)$. Then we obtain
\eq{
{\i B_2\over\pi B_1}\left.{d\over da}\left(V^{c_{-1}\overline{\tau_{-1}(h)}}_a(x)
+G_{-a}^{-1}V^{c_{-1}\overline{\tau_{-1}(h)}}_{-a}(x)\right)\right|_{a=a_{11}}
=G_{-1,1}^{-1}V^\bh_{-1,1}(x)-G_{13}^{-1}V^\bh_{13}(x).
\label{V'V'VV-id}
}
If we take $h=r_{-1,1}(\ft_{-2k})c_{-1}$, we can see the first term in the r.~h.~s.\ to be just negated the r.~h.~s.\ of~(\ref{TTheta-V-negspin}). Taking the sum of (\ref{TTheta-V-negspin}) and (\ref{V'V'VV-id}), we obtain
\Multline{
V^{c_{-1}\bar\ft_{-2k-2}}_{11}(x)
+{\i B_2\over\pi B_1}\left.{d\over da}\left(V^{c_{-1}\overline{\tau_{-1}r_{-1,1}(\ft_{-2k})c_{-1}}}_a(x)
+G_{-a}^{-1}V^{c_{-1}\overline{\tau_{-1}r_{-1,1}(\ft_{-2k})c_{-1}}}_{-a}(x)\right)\right|_{a=a_{11}}
\\
=-G_{13}^{-1}V^{\overline{r_{-1,1}(\ft_{-2k})c_{-1}}}_{13}(x).
\label{TThetaplus-11alt-posspin}
}
Comparing this with (\ref{TThetaplus-11-negspin}), we obtain
\eq{
V^{\overline{r_{11}(\ft_{-2k-2})}}_{11}(x)
=V^{\bar\ft_{-2k-2}}_{11}(x)+{\i B_2\over\pi B_1}\left.{d\over da}
V^{\overline{({\rm id}+r_a)\tau_{-1}r_{-1,1}(\ft_{-2k})c_{-1}}}_a(x)\right|_{a=a_{11}}.
\label{t2k-refl11}
}
Surely, this formula is not quite explicit, since its r.~h.~s.\ contains the reflection map. Nevertheless, the reflections in the r.~h.~s.\ are regular, so that it correctly shows us the form of the singular part. This expression becomes explicit for $k=0$, where it reduces to~(\ref{c2-refl11}), and~$k=1$, where the reflections in the r.~h.~s.\ are all known.

\section{Conclusion}

We have studied the algebra underlying the free field construction for form factors in the sinh-Gordon and related models. We found this algebra to be a subalgebra in a product of the deformed Virasoro algebra and a particular Heisenberg algebra in the limit $t\to-q$, $|q|\to1$. We have constructed the screening operators, which commute with the algebra, and studied the structure of the Fock spaces as modules of the algebra. It allowed us to find identities between form factors of the operators from different families. We have found three types of these identities: the resonance identities on even levels; the equation of motion; the conservation laws. The last two types of identities were already known, but our proof seems to be much simpler and more straightforward. It does not involve any recursion relations, which become too complicated on higher levels. The resonance identities on even levels had not been proved before and correspond to the identities discussed in~\cite{Lashkevich:2011ne}. We hope our construction to provide a basis for the proof of odd level resonance identities, which generalize the equation of motion.

A surprising feature of our construction for singular vectors is its striking resemblance to the structures appeared in the work~\cite{Babelon:1996sk} by Babelon, Bernard and Smirnov for quite a different case: the kink form factors of the sine-Gordon model at the reflectionless points. These authors introduced a kind of fermion construction for form factors of descendant operators. We think it not to be a coincidence and expect the fermion construction can be bosonized in terms of the free field representation used here.

A problem, which has not been addressed here, but have to be worked out is a description of the restricted sine-Gordon models, which are $(1,3)$ perturbations of the minimal models. While we only consider the breather form factors, it is possible to describe the perturbed models of the $(2,2N+1)$ series. The models of this series are obtained from the sine-Gordon model with $b^2=-2/(2N+1)$ by identifying the $N$th breather with the first breather~\cite{Smirnov:1990vm}. In the framework of the free field representation for form factors this condition seems to be highly nonlinear and very complicated for solving. It is necessary to elaborate a simpler condition, presumably based on the screening operators.

\section*{Acknowledgments}

We are grateful to A.~Belavin, M.~Bershtein, B.~Feigin, F.~Smirnov for many useful discussions. M.~L.\ is thankful to LPTHE, Universit\'e Pierre et Marie Curie and, especially, to F.~Smirnov for the hospitality during his visit in October 2012 and LIA Physique Th\'eorique et Mati\`ere Condens\'ee for the financial support of this visit. The study was supported, in part, by the Ministry of Education and Science of Russian Federation under the contracts \#8410 and~\#8528 and by Russian Foundation for Basic Research under the grant \#13-01-90614.

\raggedright

\end{document}